\newif \ifdraft    \draftfalse
\pdfoutput=1

\documentclass[letterpaper,twocolumn,10pt]{article}
\usepackage{usenix,epsfig,endnotes,enumitem, multicol,xcolor}
\usepackage{hyperref}
\hypersetup{hidelinks}

\usepackage{amsmath,amsfonts,amsthm,bm} \usepackage{algorithm2e}
\usepackage[numbers]{natbib}
\usepackage{cleveref}
\usepackage{subfigure}

\usepackage{authblk}

\usepackage{breakurl}
\usepackage[compact]{titlesec}
\newtheorem{definition}{Definition}

\setcitestyle{square}
\abovecaptionskip
\belowcaptionskip
\date{April 20, 2022}

\title{\Large \bf Veritas: Answering Causal Queries from Video Streaming Traces}

\begin{document}

\author[1]{Chandan Bothra$^*$}
\author[1]{Jianfei Gao$^*$}
\author[1]{Sanjay Rao}
\author[1]{Bruno Ribeiro}
\affil[1]{Purdue University}


\maketitle
\def\thefootnote{*}\footnotetext{These authors contributed equally to this work.}\def\thefootnote{\arabic{footnote}}


\newcommand{\SHOWCOMMENT}[1]{\ifdraft {[#1]}\fi}
\newcommand{\sgr}[1]{\SHOWCOMMENT{{\color{red} SGR: #1}}}
\newcommand{\bruno}[1]{\SHOWCOMMENT{{\color{blue} BRUNO: #1}}}
\newcommand{\gao}[1]{\SHOWCOMMENT{{\color{purple} Jianfei: #1}}}
\newcommand{\cb}[1]{\SHOWCOMMENT{{\color{brown} Chandan: #1}}}

\newcommand{\doop}{\text{do}}
\newcommand{\System}{Veritas}
\newcommand{\Baseline}{Baseline\xspace}
\newcommand{\GT}{GTBW\xspace}
\newcommand{\capacity}{GTBW\xspace} 
\newcommand{\Capacity}{GTBW\xspace}
\newcommand{\SystemLow}{SystemLow\xspace}
\newcommand{\SystemHigh}{SystemHigh\xspace}
\newcommand{\ModelFtxt}{$f$\xspace}
\newcommand{\ModelFmath}{f}
\newcommand{\ModelF}{\ModelFtxt}

\newcommand{\cH}{\mathcal{H}}
\newcommand{\cO}{\mathcal{O}}
\newcommand{\cC}{\mathcal{C}}
\newcommand{\cE}{\mathcal{E}}
\newcommand{\cS}{\mathcal{S}}
\newcommand{\pdo}{\texttt{do}}
\newcommand{\indep}{\bot\!\!\!\!\bot}

\SetKwComment{Comment}{/* }{ */}

\begin{abstract}
In this paper, we seek to answer \textit{what-if} questions -- i.e., given recorded data of an existing 
deployed networked system, what would be the performance impact if we changed the design of the system 
(a task also known as causal inference).
We make three contributions.
First, we expose the complexity of causal inference in the context of adaptive bit rate video streaming, a 
challenging domain where the network conditions during the session act as a sequence of latent and confounding 
variables, and a change at any point in the session has a cascading impact on the rest of the session.
Second, we present \System{}, a novel framework that tackles causal reasoning for video streaming 
without resorting to randomized trials.
Integral to \System{} is an easy to interpret domain-specific ML model (an embedded Hidden Markov Model) that relates 
the latent stochastic process (intrinsic bandwidth that the video session can achieve) to actual observations 
(download times) while exploiting control variables such as the TCP state (e.g., congestion window)
observed at the start of the download of video chunks.
We show through experiments on an emulation testbed that \System{} can answer both counterfactual 
queries (e.g., the performance of a completed video session had it
used a different buffer size) and interventional queries (e.g., estimating the download time for every possible video quality choice for the next chunk in a session in progress).
In doing so, \System{} achieves accuracy close to an ideal oracle, while significantly 
outperforming both a commonly used baseline approach, and Fugu (an off-the-shelf neural 
network) neither of which account for causal effects.
\end{abstract}


\section{Introduction}
\begin{figure}[t]
  \centering
  \includegraphics[width=0.48\textwidth]{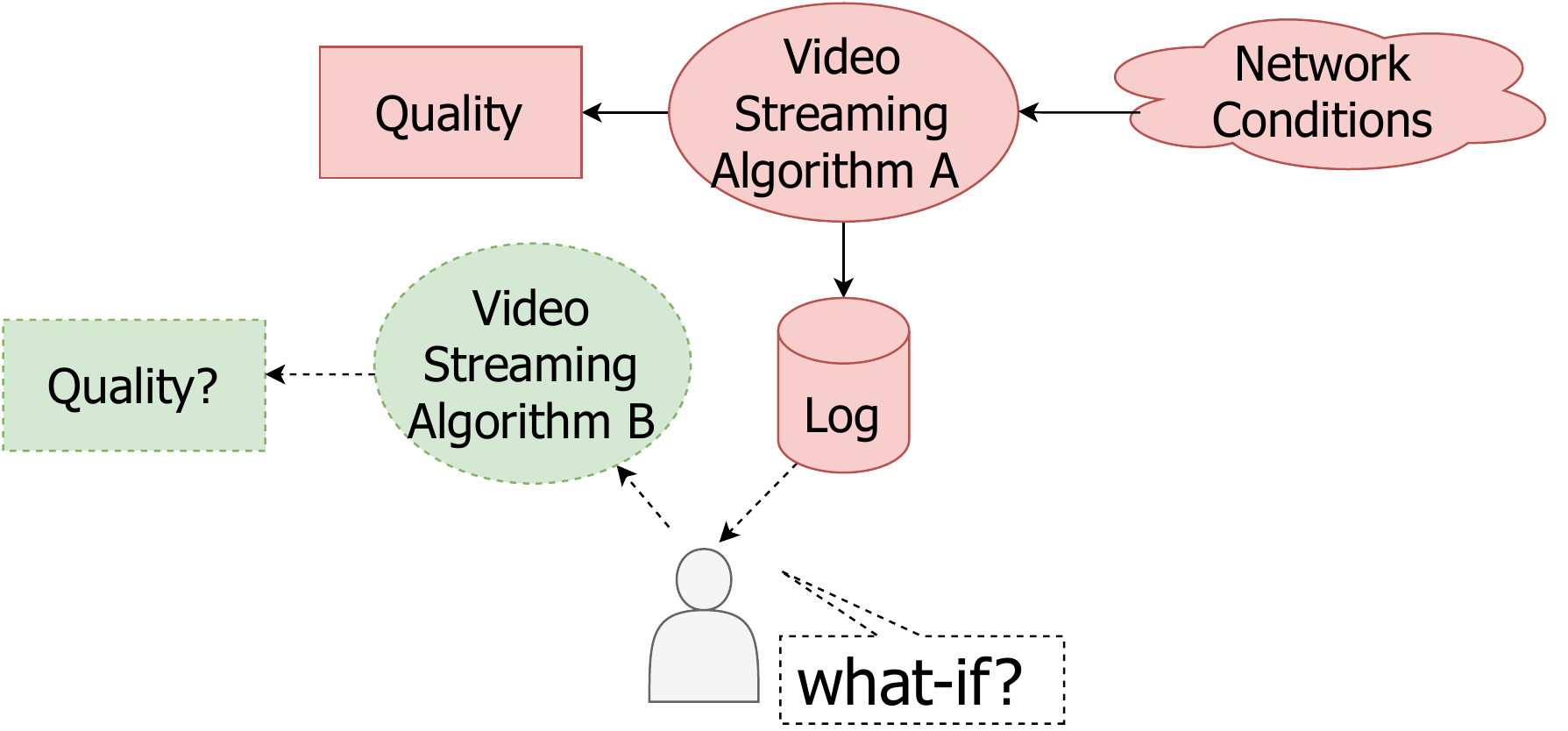}
  \caption{Example {\em what-if} question asked by a network designer: what would be the quality if algorithm B had been used instead of A under the same network conditions?}
  \label{fig:abr-what-if}
\end{figure}

A central theme of data-driven networking is answering \textit{what-if} questions --- given data obtained from a real-world deployment of an existing system, we want to infer what would have happened if we had used a different system design. For instance, given data collected from real video streaming sessions, a video publisher may wish to 
understand the performance if a different Adaptive Bitrate (ABR) algorithm were used 
(\Cref{fig:abr-what-if}), or if a new video quality (e.g., an 8K resolution) were added to the ABR selection, or an existing bit rate choice were removed (e.g., during the COVID crisis, many video publishers restricted the maximum bit rate~\cite{netflix-youtube-reduce-bitrate}). Answering \textit{what-if} questions of this nature is also known as \textit{causal reasoning}. Causal inference considers the effect of events that did not occur while the data was being recorded~\cite{pearl_causality_2009}, and has been explored in domains as diverse as economics~\cite{angrist1996identification} and epidemiology~\cite{rothman_causation_2005}.

\paragraph{Shortcomings of traditional (associational) machine learning.}
Several widely used machine learning (ML) tools are inadequate for causal inference.
Many approaches (e.g., neural networks and decision trees) merely capture correlations in collected data, limiting them to \textit{associations predictions,} i.e., predictions that are related to associations between observations in a deployed 
system. Associations, however, are inadequate to answer causal questions. For instance, people carrying umbrellas on a sunny morning
is a good predictor of rain in the afternoon. 
However, forbidding people to  carry umbrellas in the morning does not {\em prevent} rain in the afternoon. 
Similarly, in video streaming, 
an ABR algorithm could choose lower bitrates when network conditions are poor, resulting in an association between lower video bitrates and rebuffering events. However, decreasing bitrate will not {\em cause} more rebuffering events -- rather, the opposite is likely to happen.
%
%
Other approaches such as Reinforcement Learning and Randomized Control Trials allow reasoning about a
redesigned system but require \textit{active interventions} that involve changing a system, and observing 
its performance among real users. These approaches could be disruptive to the performance of real users, 
and cannot answer {\em what-if} questions about past sessions (\S\ref{sec:background}).

\paragraph{Confounders in video streaming.}
In contrast to the above approaches, our work focuses on \textit{causal} inference on 
\textit{passively collected data}, which is not disruptive to the performance of live users. 
We consider causal inference not only about how the proposed change would affect sessions in the future
(also referred to as \textit{interventional inference}) but also how it would have affected a given session in the past
(also referred to as \textit{counterfactual inference}). We expand on the distinctions in \S\ref{sec:causalQueries}.

While causal inference can benefit many networking tasks, in this work we focus on video streaming. First, it is a domain where there has been much interest in using data to drive design optimizations~\cite{xatuSigmetrics22,liu2012case,krishnan_video_2012,bartulovic_biases_2017,cs2p,yan_learning_2020}. Second, video streaming relies on adaptive bit rate (ABR) algorithms, where decisions made by the algorithm depend on network conditions, which in turn impact observable measurements. Owing to the adaptive nature, the {\em network conditions encountered  during the session} act as \textit{a sequence of latent confounding variables}, resulting in complex spurious correlations in data, which can impair the use of common ML approaches. We demonstrate this by showing that Fugu~\cite{yan_learning_2020}, a recent work that uses a neural network to predict download times in video sessions, can suffer significant biases when asking causal questions (\S\ref{sec:challenge}). 

\paragraph{Cascading effects complicate causal inference in video streaming.}
The dynamic nature of video streaming makes causal inference a challenging task. 
Consider asking the following {\em what-if} question for a recorded video session: what if bitrate $b'$ rather than the original $b$ had been chosen for video chunk $n \geq 1$, $b' \neq b$?
This {\em what-if} change (from $b$ to $b'$ at chunk $n$) has a cascading impact on the session's future buffer occupancy, and  bitrate selection decisions, as well as the start times of future chunk downloads. Thus, all observed variables describing chunk $n' \geq n$ can potentially change due to a different decision for chunk $n$. 
Here, the data recorded in the session after chunk $n$ no longer represents what will happen in the session even if no other changes were made in the future.



\paragraph{Taming the complexity of causal inference with \System{}.}
Motivated by the above challenges we design \textbf{\System{}}, a novel framework for answering causal queries for video streaming. Rather than complex ML models, or resort to randomized trials, \System{} only relies on easy-to-interpret and low-complexity ML models, while only requiring pre-recorded data. 
The challenge that \System{} tackles is {\em abduction}~\cite[Section 4.2.4]{pearl_causality_2009}, which involves (i) inferring a set of likely values for latent variables consistent with the observations; and (ii) modeling the proposed changes to return the answer to a {\em what-if} query using the inferred latent variables. While abduction is challenging in general,
the key insights of \System{} are (a) a careful selection of control variables (the TCP states at the start of each chunk download) that simplifies the causal task, and (b) a ML method to perform abduction that is principled, yet accesible and easy to interpret given it leverages domain insights.

More specifically, as part of \System{}, we have designed a domain-specific ML model that relates the latent stochastic process 
(intrinsic bandwidth that the video session can achieve if TCP were in steady state throughout the session) to actual observations (actual throughput observed by chunk downloads), when also given a sequence of additional control variables in the form of the TCP states at the start of each chunk download. This control is needed since the actual observed throughput depends on the TCP state of the connection (e.g., whether slow-start is in progress), and the size of the downloaded object.
The control allows us to ``invert'' the observed throughput variables in order to get the latent 
bandwidth variables. 

To ensure we represent the statistical dependencies in the latent bandwidth time series during the inversion process, we develop an Embedded Hidden Markov Model (EHMM), which embeds a domain-specific model for the emission process.  A Bayesian posterior sampling of the EHMM allows us to capture the uncertainty inherent in the combination of our inversion, stochastic modeling, and the data. Once a sampled inverted bandwidth process is obtained, 
we can now directly evaluate the proposed changes, and return the answer to the {\em what-if} query. Rather than a single {\em point estimate}, \System{} provides a range of potential outcomes reflecting the inherent uncertainty in inferences that can be made from the data.

\paragraph{Evaluation.}
We evaluate \System{} with respect to its ability to answer a range of {\em what-if} causal queries including the impact of (i) changing the ABR algorithm; (ii) changing the buffer size; and (iii) changing the set of video qualities that the ABR algorithm could select from using an emulation testbed.
Our evaluation approach involves emulating a video streaming system in its original setting, 
mimicking a deployed system with bandwidth traces that serve as the ground truth.
We then apply \System{} on the logs of the deployed systems (excluding the ground truth bandwidth traces) and use \System{}'s abduction to predict the impact of the {\em what-if} change.
We compare the predictions from \System{} with predictions from a baseline approach that uses the logs directly without explicit causal adjustments, and an oracle approach that knows the exact ground truth bandwidth values.
Across the board, \System{} returns significantly more accurate results to causal inferences than the baseline approach, and close to the ground truth values. For example, when changing to high video qualities, \System{} predicted negligible rebuffering ratio across all the traces, close to the oracle, 
while Baseline predicted a much higher median rebuffering ratio value of around 6.7\%.

We also evaluate \System{}'s ability to answer interventional queries (inferences related to the future) by focusing on its ability to predict the download time of future video chunks given information about past chunk download statistics. Note that \System{} must be able to make predictions under new unseen scenarios (e.g., session logs of the previously deployed ABR algorithm may only contain certain chunk size sequences, while the intervention may need to make decisions about more general sequences). We show that for such interventional queries, \System{} achieves much higher accuracies than Fugu~\cite{yan_learning_2020}, which relies on an associational method. Overall, the results show the importance and benefits of \System{}.

\section{Background and Motivation}
\label{sec:background}
In this section, we motivate the need for causal reasoning, and why ML tools used for associational predictions, and approaches such as Reinforcement Learning and Randomized Control Trials fall short. We illustrate this in the context of video streaming, and show how a state-of-the-art video streaming system~\cite{yan_learning_2020} that uses associational reasoning for a causal task falls short.

\subsection{Causal vs.\ Associational Queries}
\label{sec:causalQueries}
Video streaming today typically involves splitting video into chunks, each encoded at multiple qualities. Clients pick qualities for each chunk using Adaptive Bit Rate (ABR) algorithms so as to balance between achieving high video quality, while avoiding rebuffering based on network conditions~\cite{bb,bola,mpc,festive,tian,pensieve, oboe}.

Consider data collected from a video streaming system, where for each session information is collected regarding the chunk size and the download time. Two questions may be of interest to a designer:

\noindent
\textbf{Q1.}
Given a set of observations of chunk sizes and download times of a video session, if the video session were going to next download a chunk of size $s$, what would be the download time?

\noindent
\textbf{Q2.} Given a set of observations of chunk sizes and download times of a video session, if the designer had {\em intervened} in the session and had asked to next download a chunk of size $s'$, $s' \neq s$, what would be the download time?

Question {\bf Q1} pertains to passively observing the system at hand with its existing ABR algorithm and settings. These offline observations can be used to make predictions about the system under similar conditions. More broadly, an \textit{associational prediction} seeks to predict outcomes of a system without interfering (\textit{intervening}) with its operation. In contrast, many real-world networking tasks are like {\bf Q2}, which require going beyond passively predicting the outcomes of an existing system. These tasks require \textit{causal inference}, which predicts the outcome of an intervention, a change in the way the system operates. 
Specifically, {\bf Q2} pertains to the impact of an \textit{interventional change} to the system design: specifically using a different decision procedure that leads to a chunk of size $s'$ being downloaded rather than the original ABR's decision of downloading size $s$. More generally, the designer may wish to understand the implications on performance if some aspect of the system
were changed (e.g., changing the set of video qualities the client could choose from, the buffer size, or the ABR algorithm).

We next define the two types of causal inference algorithms of interest in our work, refining a common definition of {\em learning algorithms} \cite[Chapter 1.1]{Mitchell1997}.
\begin{definition}[Learning interventional inference for network tasks]
\label{def:intervention}
Given (i) a networking  task over an unseen session with a new method;
(ii) training experience (e.g., existing recorded sessions) obtained with an old method; and (iii) a performance measure; then,
a computer program is said to {\em learn interventions} if its prediction performance in the new method at the task (new sessions) improves if given more experience (e.g., given more recorded sessions with the old method).
\end{definition}
We further refine \Cref{def:intervention} for counterfactuals.
\begin{definition}
[Learning counterfactual inference for network tasks] 
\label{def:counterfactual}
Given (i) training experience (e.g., existing recorded sessions) running an old method; (ii) a new method; and (iii) a performance measure, a computer program is said to learn to perform {\em counterfactual inference} if its ability to predict the performance of the new method if it had been used in place of the old method in the same recorded sessions improves if given more experience (e.g., given more recorded sessions with the old method).
\end{definition}
In this work we introduce \System{}, a computer program that is able to perform both interventional and counterfactual learning (\Cref{def:intervention,def:counterfactual}, respectively).

\begin{figure*}[h]
\subcaphangtrue
\subfigure[]{
\includegraphics[height=4.25cm]{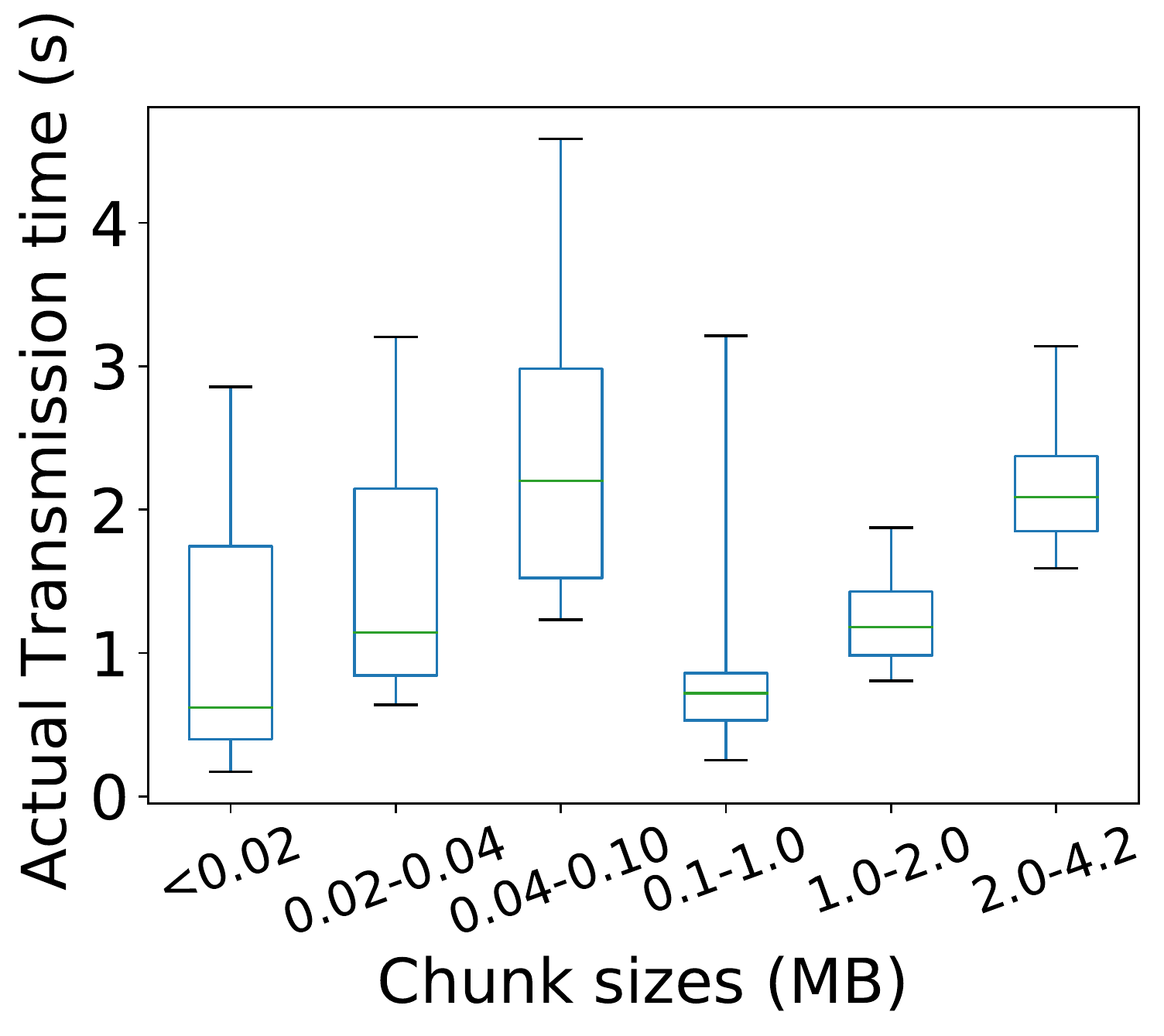}
\label{fig:exp-bias-distribution}
}
\subfigure[]{
\includegraphics[height=4.25cm]{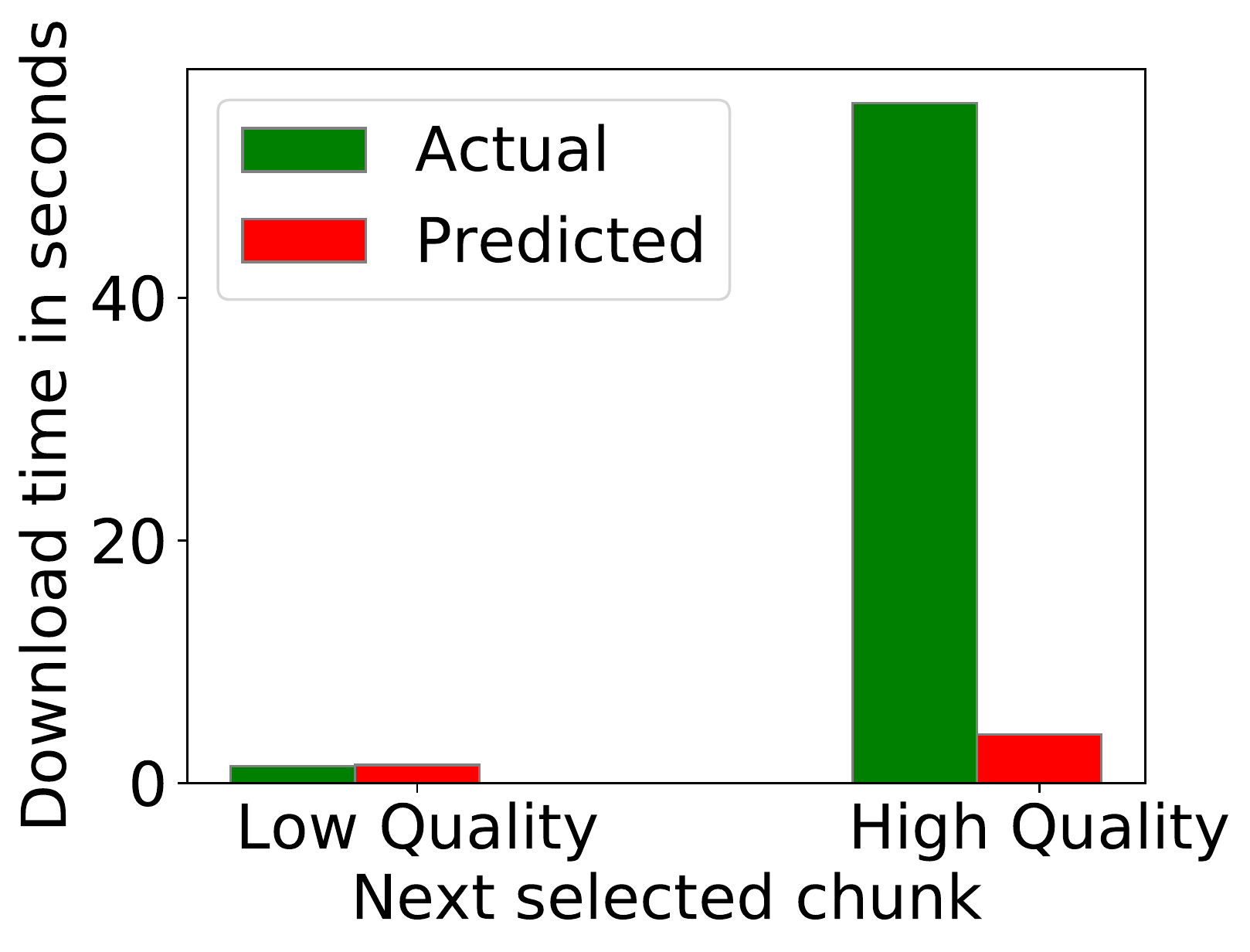}
\vspace{-2cm}
\label{fig:fugu-bias}
}
\subfigure[]{
\includegraphics[height=4cm]{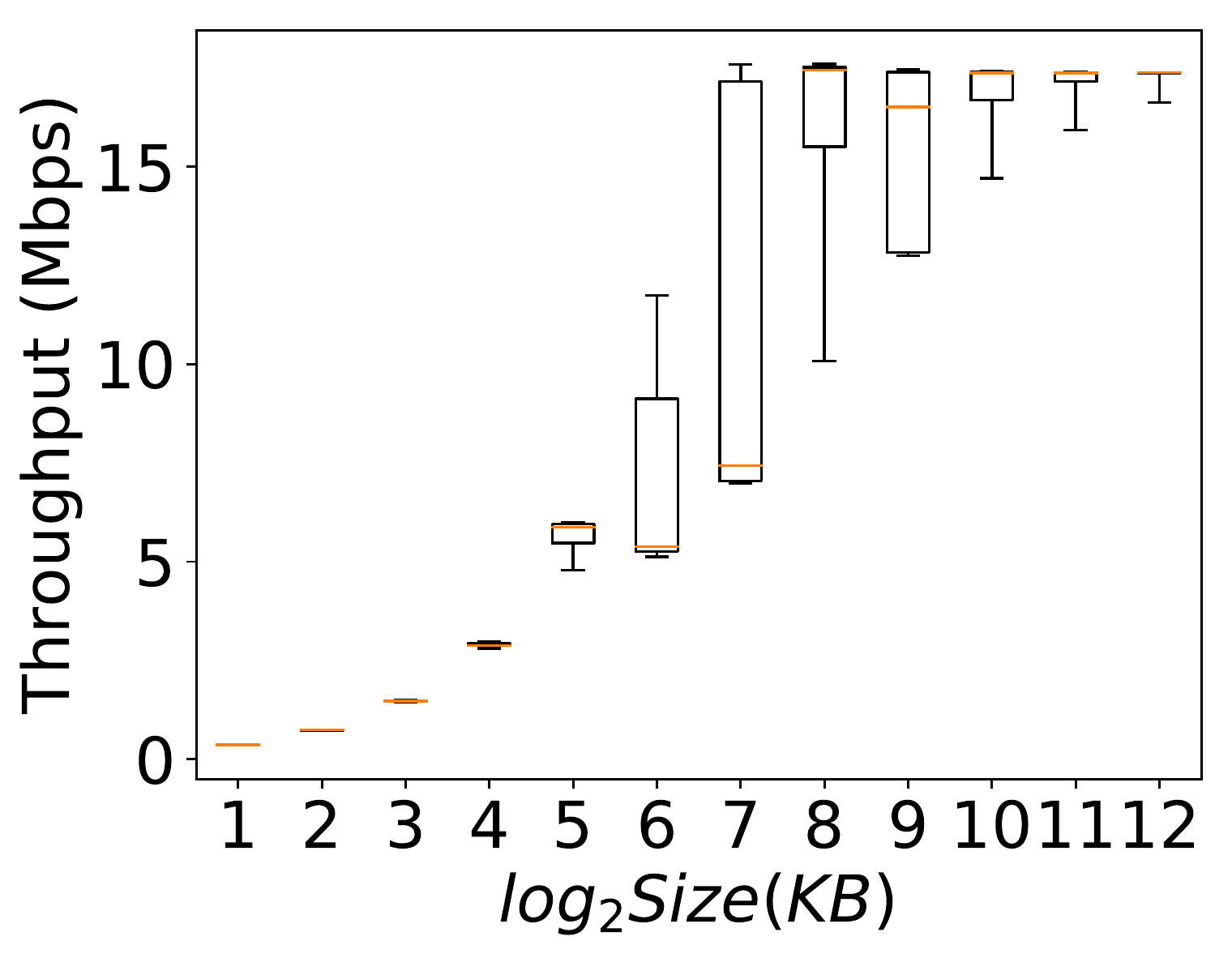}
\label{fig:throughput-variance}
}
\label{fig:motivation}
\caption{(a) Distribution of download times for different groups of chunk sizes with the MPC algorithm on a subset of FCC traces ~\cite{fcc}. The relationship is not monotonic owing to the adaptive nature of the algorithm. (b) Prediction error in Fugu ~\cite{yan_learning_2020} with causal queries. (c) Variance in observed throughput with chunk size for same emulated network bandwidth.}
\end{figure*}


\subsection{Challenges with causal queries}
\label{sec:challenge}

\textbf{Associational approaches are inadequate.}
Most ML methods work by learning associations in existing data, and, hence, are only appropriate for associational predictions. Unfortunately, the result of an associational prediction may be wildly inaccurate for a causal question. We next illustrate this in the context of Fugu~\cite{yan_learning_2020}, which uses an associational ML model for a causal query. Specifically, Fugu~\cite{yan_learning_2020} proposes a neural network which predicts the download time of a video chunk given its size, and given the size and the download times of the previous $K$ chunks. Consider Fugu trained with data obtained from the deployment of an ABR algorithm, say Algorithm A. This effectively trains Fugu to answer the associational Question \textbf{Q1} which involves predicting the download time for the chunk size selected by Algorithm A (\S\ref{sec:causalQueries}). However, consider that the trained Fugu model is actually deployed as a predictor in a real video streaming session, in a manner that {\em intervenes} with bit rate selection. That is, at any given time step of a live session, Fugu is used to predict the download times for \textit{all possible chunk sizes}, and an appropriate chunk size is selected based on these predictions. Then, effectively, Fugu is being used to tackle the causal query \textbf{Q2} (\S\ref{sec:causalQueries}). 

Unfortunately, the associational approach suffers from a bias because the deployed ABR algorithm A tends to pick lower (resp.\ higher) sized chunks when network bandwidth is bad (resp.\ good).
Consequently, the download time with an ABR algorithm many not necessarily show the expected dependence with size. To illustrate this, we conducted controlled experiments 
where we trained Fugu on 100 traces, 50 with poor network conditions [0-0.3 Mbps] and 50 with good network condition [9-10 Mbps] with the MPC algorithm in an emulation testbed (details in \S\ref{sec:evaluation-setup}) using the FCC throughput traces~\cite{fcc} \sgr{Be vaguer?}
Figure~\ref{fig:exp-bias-distribution} presents the download time of all chunks across all the video sessions, with each boxplot corresponding to chunks with a particular size range. 
The figure shows that the download times do not grow in a linear fashion with chunk size, rather show a non-monotonic dependence. This is because of the adaptive bit rate selection described above. 
  

We next test Fugu on a new trace with poor network conditions. We consider a point in time where the ABR algorithm has picked a sequence of lower quality chunks. We then use Fugu to answer the {\em what-if} questions, what would the download time be if (i) the next chunk selected were high quality; and (ii) the next chunk selected were low quality. Figure~\ref{fig:fugu-bias} shows the download times predicted by Fugu and the actual download times in each case. The figure shows that Fugu significantly underestimates download times for the high quality chunk, but does a good job for the low quality chunk. This is because Fugu uses an associational model that is effective at predicting download times for a chunk size that the deployed algorithm would have selected next, but not the download times if the chunk size had been forced to be a particular value. Consequently, while the model effectively predicts the download time of chunk sizes actually chosen by the deployed ABR, it performs poorly when answering the causal query {\em what would happen if an alternate size were chosen}. 



\begin{figure*}[t]
        \includegraphics[width=0.98\textwidth]{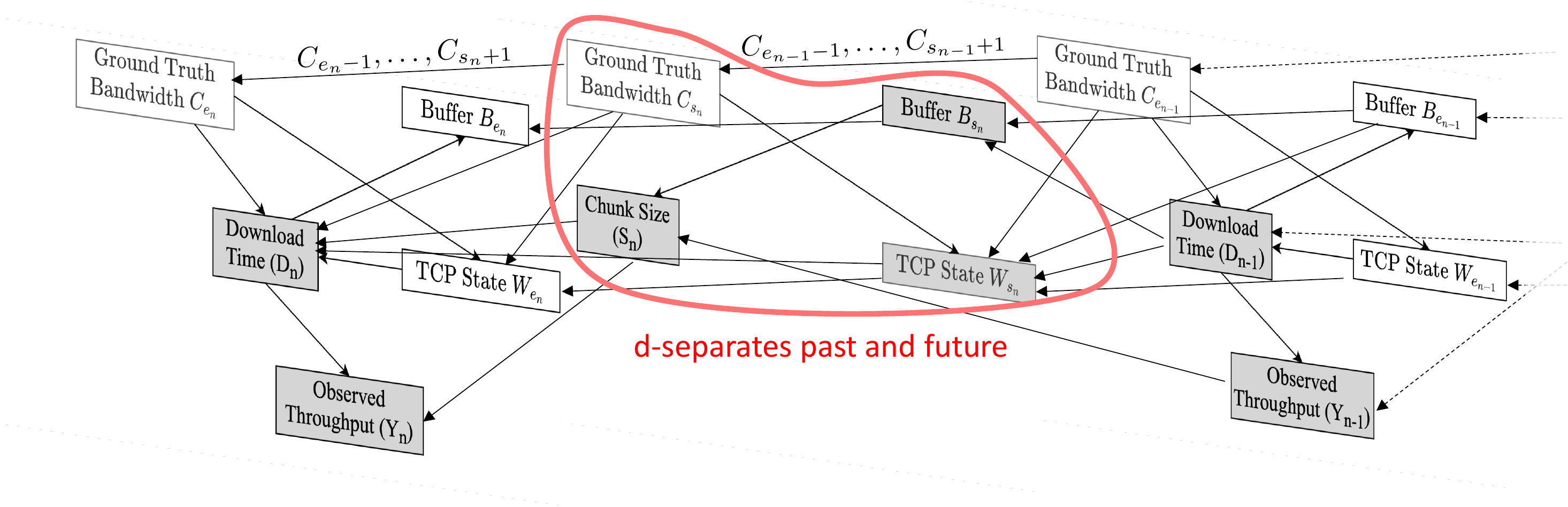}
        \vspace{-15pt}
        \caption{Causal model of (embedded) dependencies in an ABR algorithm starting at $e_{n-1}$, the arrival time of the $(n-1)$-st chunk, until $e_{n}$, the arrival time of the $n$-th chunk. Shaded gray variables are observed, while white (unshaded) variables are hidden. A sufficient condition for a set $U$ of variables to $d$-separate a set $A$ and $B$ is that all undirected paths in the DAG between $A$ and $B$ include at least one variable from $U$, and no such paths have arrows collide ``head-to-head'' in the variables in $U$.}
        \label{fig:design-cdg}
\end{figure*}

\noindent
\textbf{Latent variables complicate causal queries.}
A potential alternative to the above approach is to consider the observed throughput when downloading a video chunk, defined as
$S/D$. where $S$ is the size, and $D$ the download time.
For instance, the neural network from~\cite{yan_learning_2020} could use a sequence of observed chunk throughputs, and predict the throughput of the next chunk. Unfortunately, the observed throughput itself is dependent on the size of chunks owing to TCP slow start effects~\cite{yan_learning_2020,
bartulovic_biases_2017,xatuSigmetrics22}.
Figure~\ref{fig:throughput-variance} presents a distribution of throughput for chunks in a given size range in controlled experiments using TCP where we emulated a constant network bandwidth
of 18 Mbps in a client server setup, and sent payloads of varying sizes (2KB to 4MB). \sgr{Did we also create random gaps?} \cb{Yes}Note that the gap between payloads impacts whether TCP enters a slow-start restart phase~\cite{ssr}. The graph shows that for small sizes (less than the bandwidth delay product of the network), throughput is much smaller, while it is closer to the intrinsic network bandwidth for larger sizes. Note that for intermediate sizes (around $2^{6}$ to $2^{10}$ Kilobytes) there is significant variability in throughput based on the gap between the transmission of two consecutive payloads, and whether TCP enters slow start restart or not. Thus, simply considering throughput is insufficient, and while it would be desirable to consider the intrinsinc network bandwidth, this is a hidden variable not available in data. 

  

\subsection{Why not actively intervene?}
\label{sec:RCT}
Rather than making predictions by passively observing a system, some approaches can evaluate the impact of a design change by actively intervening (changing) the system, and observing the performance. These techniques include Randomized Control Trials (RCTs), A/B Testing, and Reinforcement Learning~\cite{Sutton2002}.

\textbf{RCTs~}\cite{deaton_understanding_2018} can measure the effects of interventions. However, there are several disadvantages to RCTs. First, such trials may lead to degraded performance to some viewers. For instance, in the context of video streaming algorithm, even if chunk bitrates
are chosen randomly without regard to network conditions, a chunk's start time is not random as it still depends on the download end time of the previous chunk~\cite{sruthi2020netai}. 
Another approach is to use \textbf{A/B testing} of a new design change. Since A/B testing can impact the performance of live users, it is only used in a conservative fashion if an offline analysis approach indicates the design change has sufficient potential. Thus, there remains a need to answer a question offline with trace data alone.  

\textbf{Reinforcement learning} may be viewed as a sequential RCT in that the agent dynamically learns the best decisions to take at each state of the system. The collection of all pairs (best decisions, current system states) is denoted a {\em policy}. A drawback of RCTs in general, and reinforcement learning in particular, is that it only answers the question of which decisions, out of a set of tested decisions, are the best to take for a given system state. If our set of possible decisions changes, the RCT/RL tests must be run again on new sessions.

Another important point is that both RCTs and reinforcement learning cannot directly answer counterfactual queries, although their randomized (exploration) measurements may still be used by counterfactual estimators in some special cases (e.g.,~\cite{bareinboim2015bandits}).
Hence, interventional methods may not be useful in some scenarios because they can only be tested in future sessions: Imagine seeing rare network conditions where a deployed algorithm performed poorly. Since these conditions are rare, we would like to know if a certain change to the algorithm would have significantly improved the performance. This is a counterfactual query and RCTs and reinforcement learning are generally not applicable in this scenario since the event is in the past, and any RCT to test a new intervention on the system can only be applied in future sessions.

\section{\System{}: A causal inference framework for video streaming}
\label{sec:causalFramework}
In this section, we present \System{}, our framework for answering causal queries related to video streaming.
We start by presenting a causal graph (DAG) which models the variables involved with video streaming, and the dependencies or causal relationships between them (\S\ref{sec:causalDAG}). We then discuss how the DAG leads us to decide what adjustments are needed to identify a causal effect, and 
how to make the adjustment (\S\ref{sec:abduction}).
Finally, \S\ref{sec:design:causalQueries} discusses how \System{} puts all these methods together to perform counterfactual and interventional inference.

\subsection{Modeling causal dependencies}
\label{sec:causalDAG}
A key factor that impacts the decisions made by a video streaming algorithm is the {\bf Ground Truth Bandwidth (henceforth abbreviated as \Capacity{}}), which captures the bandwidth the network is intrinsically capable of, without considering dependence on size, and the slow start effects of the transport protocol -- i.e., what the transport protocol would intrinsically see if it were running in steady state. We model the evolution of \Capacity{} as a discrete process over discrete time intervals $t \in \{1,\ldots,T\}$ (each of wall-clock time length of $\delta$), with the \capacity during any time interval being a constant. Time is assumed to be discrete to simplify our approach, since $\delta$ can be as fine-grained as necessary.

Consider that the session downloads a series of chunks $1 \dots N$. Chunk $n \in \{1,\ldots,N\}$ starts its download at time $s_n \in \{1,\ldots,T\}$ and finishes at time $e_n \in \{s_n,\ldots,T\}$. \gao{use interval instead of time?}
The variables that evolve over time are: (i) $C_t \in \cC$, the average \Capacity{} at time interval $((t-1)\delta,t\delta]$; (ii)  $B_{t},$ the amount of buffer in the video player at time $t \in \{1,\ldots,T\}$, and (iii) $W_{t},$ the TCP state at time $t$. The TCP state includes parameters such as the congestion window, slow start threshold, RTT, min RTT, time since last data send, and RTO. 

The variables that evolve at each chunk request are: (i) the size ($S_n$) of the $n$-th requested chunk and (ii) $D_n,$ its download time, $n=1,\ldots,N$. The throughput observed during the download ($Y_n$) can be calculated using the chunk size and download time. 

Henceforth, for any random variable $X$ we define the sequences $X_{a:b}:=(X_a,\ldots, X_b)$ and $X_{s_{a:b}}:=(X_{s_a},\ldots,X_{s_b})$.
Moreover, 
let $\cS = \cup_{n =1}^N \{s_n\}$ and $\cE = \cup_{n =1}^N \{e_n\}$ be the set of random variables of showing the discrete times where a chunk starts and ends downloading, respectively.
We assume that the variables in $W_{s_{1:N}}$, $B_{s_{1:N}}$, $S_{1:N}$, $\cS$, $\cE$ and $Y_{1:N}$ (shown in shaded gray in \Cref{fig:design-cdg})
are generally \textit{observed} variables in video streaming sessions (that is, all the information regarding them is either directly available, or can be calculated from the data). Note that TCP state information is easy to collect (e.g,  using the \emph{tcp\_info} structure in Linux systems~\cite{tcpinfo}).
Further, although we could collect the information, we do not require the values $\{W_{t}\}_{t \in \{1,\ldots,T\} \backslash \cS}$, $\{B_{t}\}_{t \in \{1,\ldots,T\} \backslash \cS}$, and treat these variables as hidden.

\Cref{fig:design-cdg} shows a directed acyclic graph (DAG) describing the causal dependencies for video streaming. Note that \Cref{fig:design-cdg} only illustrates the embedded process of $\{C_{t}\}_{t \in \cS \cup \cE}$, $\{W_{t}\}_{t \in  \cS \cup \cE}$, and $\{B_{t}\}_{t \in \cS \cup \cE}$, at the event times where a new chunk is requested or finishes downloading.
It is important to note that the variables $C_{1:T},W_{1:T},B_{1:T}$ also evolve in the time between these chunk events, but for any time $t \in \{1,\ldots,T\} \backslash \{\cS\cup \cE\}$ that happens between chunk start and end times, the random variables $B_t$ depends only on $B_{t-1}$ (just the video being played) and $C_t$ depends only on $C_{t-1}$, but $W_t$  depends on both $W_{t-1}$ and $C_{t-1}$ if there is an active chunk download at time $t$ (and only on $W_{t-1}$ if there is no active download).

The $n$-th chunk size $S_n$ is influenced (through the ABR algorithm) by both the buffer state $B_{s_n}$ at the start of download of chunk $n$ and the last observed throughput $Y_{n-1}$ (and possibly $Y_{n-2},\ldots,Y_1$ (not shown in the DAG)).
The chunk size value $S_n$ influences the download time $D_n$. 
Further, the TCP state $W_{s_n}$ determines whether the TCP connection of the chunk download experiences slow start restart, initial congestion window etc., all of which together with $S_n$ and $C_{s_n},\ldots,C_{e_n}$ also influence the download time $D_n$. A key parameter of the TCP state is the gap since the last packet was
transmitted. This in turn depends on the video application. When the buffer is full, the player does not send further requests, but when not full, it may trigger requests immediately. Hence, $W_{s_n}$ itself depends on $B_{e_{n-1}}$,
which defines a new chunk request at time $s_n$. Finally, as discussed above $S_n$ and $D_n$ together determine $Y_n$.


\noindent
\textbf{Confounders:}
The DAG in \Cref{fig:design-cdg} shows that $C_{1:T}$ are confounder variables between $S_{1:N}$, $D_{1:N}$, and $W_{s_{1:N}}$.
Confounders are hidden variables (not available in the data) that jointly influence multiple observed variables.
Moreover, we make the simplifying assumption that $C_{1:T}$ are not influenced by any other variable in the model (that is, chunk downloads do not impact the \capacity{}). 
Our
model also assumes we are running a particular version of TCP (e.g., Cubic, or BBR) and cannot directly be used to model the impact of {\em what-if} questions where the TCP version itself might change. This would
require modeling more intrinsic hidden factors such as the number of simultaneous flows in routers, etc..

\subsection{\System{} abduction for causal queries}
\label{sec:abduction}
Since no other variables affect the confounder variables $C_{1:T}$ but $C_{1:T}$ directly or indirectly affect all other variables (i.e., all other variables are descendants of some variable in $C_{1:T}$), if we could infer $C_{1:T}$ we would be able to handle any counterfactual or interventional query needed.
This procedure to infer a confounder ($C_{1:T}$) to respond to causal queries is known as \emph{abduction}~\cite[Section 4.2.4]{pearl_causality_2009}.
Abduction involves (i) ``inverting'' the observed variables to get the hidden confounders; and (ii) then modeling the proposed changes (assuming the hidden confounder values are now known) to return the answer to the {\em what-if} query. 
Abduction approaches in the ML literature
typically rely on composable statistical models using high-level programming  languages~\citep{probtorch,stan,pyro} that, unfortunately, do not effectively deal with the use of \verb|if| statements and other deterministic decision functions common in networking.
%
%
Hence, our work proposes the \System{} custom abduction method tailored to our task, described in the next few paragraphs.

\noindent
{\bf The \System{} abduction of $C_{1:T}$.}
In our setting, abduction requires sampling the network \capacity \gao{previous is ``throughput'' which I think is wrong} given all the observations in a session: 
\begin{equation}
\label{eq:C1T}
C_{1:T} \sim P\left( c_{1:N}|S_{1:N}, D_{1:N}, B_{s_{1:N}}, W_{s_{1:N}} \right),
\end{equation}
where $R \sim g$ denotes that random variable $R$ is sampled with distribution $g$.
Once the confounding variables $C_{1:T}$
are sampled given the observed variables, we can simulate the effect of the causal query in the sampled 
$C_{1:T}$ (now assumed known).
The sampling in \Cref{eq:C1T} accounts for the non-unique nature of the ``inversion'', imposing a distribution over the {\em what-if} query results.
Obtaining $C_{1:T}$ requires connecting it to the observed variables,
including the observed throughput $Y_{1:N}$, TCP states $W_{s_{1:N}}$, a sequence of buffer states $B_{s_{1:N}}$,
and chunk sizes $S_{1:N}$. We discuss how \System{} achieves this next.

\noindent
{\bf Abduction of $C_{1:T}$ via EHMMs.}
%
\begin{figure*}[ht]
\includegraphics[width=\linewidth,height=1.4in]{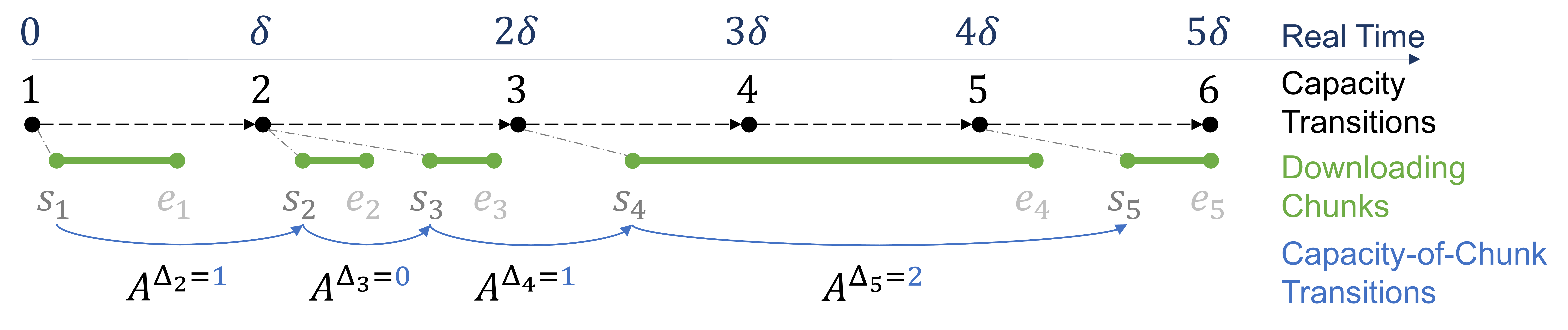}
\caption{
    {\bf Translation from Real Time Capacity Transition to Downloading Chunk Capacity Transition.}
    On the top,
    the \capacity evolves every $\delta$ time units. We assume $C_{t}$ is constant in the interval $[(t - 1)\delta, t\delta)$. In the middle lines shows five downloading chunks, beginning at $s_{n}$ and ending at $e_{n}$, $n=1,\ldots,5$.
    At the bottom, under the arrows, we show the number of \capacity transitions $\Delta_{n}$ between consecutive chunks.
    For example, chunk 2 and chunk 3 start at the same time window, so $\Delta_3 = 0$, while chunk 4 and chunk 5 both start at window 3 and 5 respectively, so $\Delta_5 = 2$.
}
\label{fig:chunk_to_hmm}
\end{figure*}
%
%
%
%
%
\System{} uses a special type of Hidden Markov Model (HMM)~\cite{viterbi2006personal}.
%
An HMM is specified by (i) a set of hidden states $\cC$; (ii) a matrix that captures the transition probabilities from one hidden state to another; (iii) a set of observations; (iv) a set of probabilities (a.k.a.\ emission probabilities), which capture the likelihood of a particular observation being generated from a given hidden state; and (v) an initial probability distribution over states. In our context, the \capacity sequence $C_{1:T} \in \cC^T$ corresponds to the hidden states, while the throughput $Y_{n}$ is the observation {\em jointly  emitted} by the states $C_{s_n:e_n}$. 

\noindent
{\bf The importance of observing TCP states $W_{s_{1:N}}$.}
Note that in an HMM, because of the Markov property, given the hidden variable ($C_{s_n}$), the emissions ($Y_n$) are independent of the past emissions ($Y_{n-1},\ldots,Y_1$).
In graphical model language we say $C_{s_n}$ needs to d-separate $Y_n$ and $\{Y_{n-1},\ldots,Y_1\}$.
A sufficient condition for a set $U$ of variables to $d$-separate a set $A$ and $B$ is that all undirected paths in the DAG between $A$ and $B$ include at least one variable from $U$, and no such paths have arrows collide ``head-to-head'' in the variables in $U$~\cite[Definition 1.2.3]{pearl_causality_2009}.
The challenge is that $C_{s_n}$ {\em does not d-separate} $Y_n$ and $Y_{n-1},\ldots,Y_1$ in the DAG of \Cref{fig:design-cdg}.

In order to achieve this d-separation independence, \System{} conditions the entire HMM on the sequence of TCP and buffer\footnote{It is in fact not necessary to observe $B_{s_{1:N}}$, as discussed in the Appendix.}  states at the chunk start times ($W_{s_{1:N}}$ and $B_{s_{1:N}}$), together with the sequence of requested chunk sizes ($S_{1:N}$).
This conditioning ultimately allows us to {\em d-separate} $Y_n$ and $\{Y_{n-1},\ldots,Y_1\}$ in the DAG of \Cref{fig:design-cdg}.
It is easy to check that the variables $C_{s_n}, W_{s_n}, B_{s_n}, S_n$ ---circled in red in \Cref{fig:design-cdg}--- block any undirected paths between $Y_n$ and $\{Y_{n-1},\ldots,Y_1\}$ in the DAG. 


\noindent
{\bf The \System{} embedded Markov chain.}
The HMM that \System{} uses (which we refer to as EHMM) also departs from standard HMM models in other ways. First, HMMs traditionally use common parameterized probability distributions (e.g., multinomial, Guassian) to model emission probabilities. Instead, \System{} embeds  a domain-specific model for its emissions. The model captures how \Capacity{}, chunk sizes, and TCP states gets translated into observed throughput.
Second, in traditional HMMs, each hidden state is associated with a single observation. However, in our context, observations are only associated with those hidden \Capacity{} states where chunks are being downloaded. The hidden \capacity itself changes during the off periods where no chunks are being downloaded, and there are no observations available during this time. Further, it is possible that there are multiple chunks downloaded in the same time interval $((t-1)\delta,t\delta]$, $t \in \{1,\ldots,T\}$. To handle this, \System{}'s EHMM allows each \capacity state to be associated with zero, one or more observations (corresponding to the number of chunks downloaded in the corresponding interval). Note that \System{}'s use of EHMM is consistent with prior work~\cite{cs2p,oboe}, which has modeled TCP throughput evolution as a Markov process, but \System{} addresses complexities associated with embedding a custom emission process, $d$-separation, and the focus is on abduction for causal inference.



\noindent
\paragraph{Hidden state transitions.}
As discussed in \S\ref{sec:causalDAG}, evolution of \Capacity{} is modeled as a discrete sequence $C_{1:T}$, where $C_t$ denotes the average \Capacity{} during time interval $((t - 1)\delta,t\delta]$. 
Further, for simplicity, \System{} uses a quantized set of capacities to ensure the number of states is discrete. \capacity{} values $\mathcal{C}$ are quantized via a hyperparameter $\epsilon > 0$.
For instance, $\epsilon = 0.5$ implies that the hidden states are $\cC = \{0.0 \text{Mbps}, 0.5 \text{Mbps}, 1.0 \text{Mbps}, \cdots\}$. 
Both hyperparameters $\delta$ and $\epsilon$ may be kept as small as needed.

The sequence $C_{1:T}$ is modeled as a first-order Markov chain
$P(C_{t} | C_{1}, \cdots, C_{t - 1}) = P(C_{t} | C_{t - 1}), 1 < t \leq T$.
The conditional distribution $P(C_{t} | C_{t - 1})$ is parameterized by a transition matrix $A$ such that
%
%
\begin{align}
\label{eqn:transmat}
    A_{i, j} = P(C_{t} = j\epsilon | C_{t - 1} = i\epsilon), \quad 1 < t \leq T.
\end{align}
For $t = 1$, since there are no transitions, we will directly model the initial distribution of \capacity by a distribution $u$, with hyperparameter
$
    u_{i} = P(C_{1} = i\epsilon).
$
Finally, given two hyperparameters, \capacity transition interval size $\delta$ and minimum \capacity discrepancy $\epsilon$, we can model the \capacity evolution by the transition matrix $A$ (\Cref{eqn:transmat}) and the initial distribution $u$, thus can measure \capacity evolution distribution by
$
    P\big( C_{1:t} \big) = P\big( C_{1} \big) \prod_{t' = 2}^{t} P\big( C_{t'} \big| C_{t' - 1} \big)
$.

\noindent
\paragraph{Domain-specific model for emission process.}
%
As discussed in \Cref{sec:causalDAG}, for any chunk $1 \leq n \leq N$, with start time $s_{n}$ and end time $e_{n}$, we will observe corresponding throughput $Y_{n}$, TCP state $W_{s_n}$ and chunk size $S_{n}$.
The throughput $Y_n$ observed by video chunk $n$ is a function of \Capacity{} $C_{s_n:e_n}$, the starting TCP state $W_{s_n}$, and the chunk size $S_n$, and we want to test if an arbitrary \capacity can fit the observed chunk.

We develop a simple model of TCP to estimate $Y_n$, denoted by~\ModelF (\Cref{alg:model_f} in the Appendix), which models congestion control with slow start, congestion avoidance and slow start restart (TCP SSR)~\cite{ssr}. If the network is idle and $W_{s_n}^{\text{last\_send}}$ is greater than the retransmission time out $W_{s_n}^{\text{rto}}$, TCP SSR is triggered and the congestion window $W_{s_n}^{\text{cwnd}}$ and slow start threshold $W_{s_n}^{\text{ssthresh}}$ are updated according to the Linux kernel implementation based on~\cite{ssr-algo}. We calculate the number of transmission rounds needed to transmit a chunk with size $S_n$ based on the updated $W_{s_n}^{\text{cwnd}}$ and $W_{s_n}^{\text{ssthresh}}$. This calculation assumes that in each round, the total data transmitted is the minimum of the Bandwidth Delay Product (BDP) of the network, or the congestion window, whichever is lower. We  model evolution of the congestion window within rounds using typical TCP slow start behavior, and using a simple additive scheme for congestion avoidance. Further, loss events are not modeled.
The number of transmission rounds, the minimum RTT in $W_{s_n}$, and the chunk size
$S_n$, are all used together to estimate the observed throughput of the video chunk. 
We emphasize that more detailed models that capture intricate details of specific TCP versions can be easily incorporated in \System{} in the future, but the above model, while simple, helps to illustrate the feasibility and potential of \System{}'s overall approach in tackling causal inference.

\begin{figure}[t]
    \centering
    \includegraphics[width=5cm]{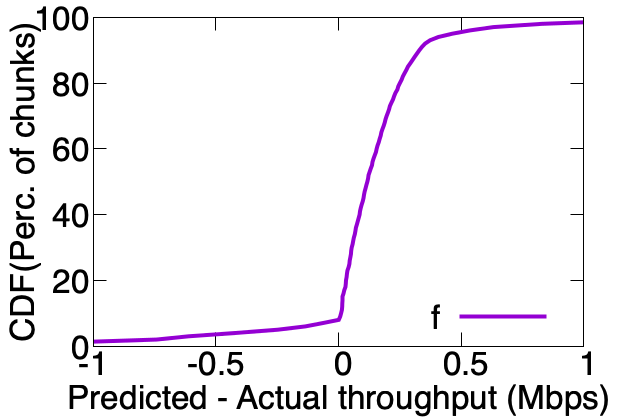}
    \caption{Relative error of~\ModelF{} shows acceptable uncertainty.} 
    \label{fig:f_accuracy}
\end{figure}

We test the performance of \ModelF in an emulation environment with a server transmitting payloads of different sizes [2KB to 4 MB] to a client, with random intervals of wait time [0.12s to 8s] between transmission of successive payloads. \ModelF derives $W_{s_n}$ using the socket stats utility in Linux~\cite{ss}. The~\Capacity{} between client and server is varied from 0.5 to 10 Mbps, and the end to end delay is varied from 5 to 40 ms using mahimahi~\cite{mahimahi} across experiments. The \Capacity{} and delay is kept constant for a particular experiment. \Cref{fig:f_accuracy} shows a CDF of the relative error between the actual throughput observed by a payload and the throughput estimated by~\ModelF across all \Capacity{} and delays. In most cases, the predicted throughput is within a range of 1 Mbps of the observed throughput by the payload.

If~\ModelF were a perfect estimator, we may have modeled emission probabilities as   $P\big( Y_{n} \big| W_{s_n}, S_{n}, C_{s_n} = c \big) = 1$ if $Y_{n} = f\big( c, W_{s_n}, S_{n} \big)$, and $P\big( Y_{n}  \big|  W_{s_n}, S_{n}, C_{s_n} = c \big) = 0$ otherwise.
To take the uncertainty of~\ModelF (\Cref{fig:f_accuracy}) into consideration, we include a white-noise Gaussian-distributed error with variance $\sigma^2$ (a hyperparameter):
\begin{align}
\label{eqn:eval_chunk_cap}
   \!\! P\left( Y_{n}  \middle|  W_{s_n}, S_{n}, C_{s_n}\!=\!c \right)\! =\! \text{Normal}\left( \ModelFmath\left( c, W_{s_n}, S_{n} \right), \sigma^2 \right)\!.
\end{align}
Note that the emission in \Cref{eqn:eval_chunk_cap} does not account for $C_{s_n+1},\ldots,C_{e_n}$.
In practice, this simplification does not have a significant impact in our ability to sample $C_{1:T}$ as our evaluation shows.
Also note that $Y_n$ in the DAG of \Cref{fig:design-cdg} does not depend on $B_{s_n},B_{s_{n-1}},\ldots$ given $W_{s_n}, S_{n}$, and $C_{s_n}$ (and $s_n$), thus there is no need to use $B_{s_n},B_{s_{n-1}},\ldots$ in estimator $f$.


\paragraph{Evolution of the embedded \Capacity.}
We next discuss how we deal with the fact that there may be no observations attached to a particular hidden state $C_t$, while there may be more than one observation for a different hidden $C_{t'}$, $t,t' \in \{1,\ldots,T\}$. 
We handle this through embedded transitions in $C_{s_{1:N}}$ in a procedure similar to Neal et al.~\cite{neal2003inferring} .
That is, for $t \in \{1,\ldots,T\}$, instead of modeling $P\big( C_{t} \big| C_{t - 1}\big)$, we model the transitions $P\big( C_{s_n} \big| C_{s_{n - 1}} \big)$, where $1 < n \leq N$. 
For chunks $n - 1$ and $n$, we will define $P\big( C_{s_n} = j\epsilon \big| C_{s_{n - 1}} = i\epsilon \big) = (A^{\Delta_{n}})_{i,j}$, where  $\Delta_{n} =  s_{n} - s_{n - 1}$ and $A$ is as defined in \Cref{eqn:transmat}.

\noindent
\paragraph{Sampling $C_{1:T}$.}
We are now ready to sample from \Cref{eq:C1T} with our \System{} EHMM.
For ease of notation, we define $I_{1:N} = (I_{1}, \cdots, I_{N})$, where $I_{n}$ is the discretized capacity index of chunk $n$, that is, $C_{s_{n}} = I_{n} \epsilon$.
Then, the maximum likelihood capacity assignment for all chunks will be
\begin{align}
\label{eqn:capbest}
    I^{\star}_{1:N} = \mathop{\arg\max}\limits_{I_{1:N}} \log\, P\big( I_{1:N} | Y_{1:N}, W_{s_{1:N}}, S_{1:N} \big),
\end{align}
where
\begin{align}
\label{eqn:capscore}
    & P(I_{1:N} | Y_{1:N}, W_{s_{1:N}}, S_{1:N})
    =  P\left( C_{s_1}\!\! =\! I_1 \epsilon \right)\\ 
    \nonumber
   &\times  P\left( Y_{1},| W_{s_1}, S_{1}, C_{s_1}\!=\!I_{1}\epsilon  \right)  \prod\limits_{n = 2}^{N} A_{I_{n - 1}, I_{n}}^{\Delta_{n}} P\left( Y_{n} | W_{s_n}, S_{n}, C_{s_n} = I_{n}\epsilon \right)
\end{align}
and $
    P\left( Y_{n}| W_{s_n}, S_{n}, C_{s_n} = I_{n} \epsilon \right)$
as defined in \Cref{eqn:eval_chunk_cap}.

To get \Cref{eqn:capbest}, we use the Viterbi algorithm which searches ---via dynamic programming--- for the values of $C_{1:T}$ that give the highest likelihood in \Cref{eqn:capscore}.
The vanilla Viterbi algorithm~\cite{viterbi2006personal} assumes a constant transition matrix, which we replace by $A^{\Delta_n}$ in \System{}.
More details of the \System{} Viterbi variant is provided in \Cref{alg:viterbi} in the Appendix.


We will then use Viterbi output (maximum likelihood estimate) to sample hidden states according to the posterior in \Cref{eq:C1T}, similar to~\cite{chib1996calculating,scott1999bayesian,robert1998reparameterization}.
For the sampling, we will additionally require the probability $P\big( C_{s_{n}} = i\epsilon, C_{n + 1} = j\epsilon \big| Y_{1:N}, W_{s_{1:N}}, S_{1:N} \big)$ which can be obtained from our variant of the Baum-Welch forward-backward algorithm (see \Cref{alg:forward_backward} in the Appendix).
We denote this pair distribution
\begin{align}
\label{eqn:joint}
    \Gamma_{i, j, n} = P\left( C_{s_{n}} = i\epsilon, C_{s_{n + 1}} = j\epsilon \middle| Y_{1:N}, W_{s_{1:N}}, S_{1:N} \right).
\end{align}
The sampling algorithm for $C_{s_{1:N}}$ is defined as \Cref{alg:sampler}.
The intermediate values $C_t$ where $t \in \cup_{n=2}^N\{{s_{n-1}+1},s_{n}-1\}$ are interpolated from sampled $C_{s_{1:N}}$.

\begin{algorithm}
\caption{{\bf Capacity Sampler.} It obtains the last state $N$ as the last state of Viterbi output, then forward samples each state $1 \leq n < N$ based on sampled state $n + 1$ and scores defined by \Cref{eqn:joint}.}
\label{alg:sampler}
\KwIn{State space $\mathcal{C}$, Length $T$, Viterbi output $I^\star_{1:N}$, Transition $A$, Pair distribution $\Gamma$}
\KwOut{A sampled capacity trace $C$}
$C_{s_N} = I^\star_N \epsilon$ \\
\For{$n = N - 1 \text{\bf ~to } 1$}{
    $\xi_{n, i} = \Gamma_{i, C_{s_{n + 1}}, n + 1}$,  $i\epsilon \in \cC$ \\
    $Z_{n} = \sum_{i \in \mathcal{C}} \xi_i$ \\
    $\pi_{n, i} = \xi_{n, i} / Z_n$, $i\epsilon \in \cC$ \\
    $C_{s_n} \sim \text{Multinomial}(\pi_{n, :})$ \\   
}
\end{algorithm}

\begin{figure}[t]
    \centering
    \includegraphics[width=5cm]{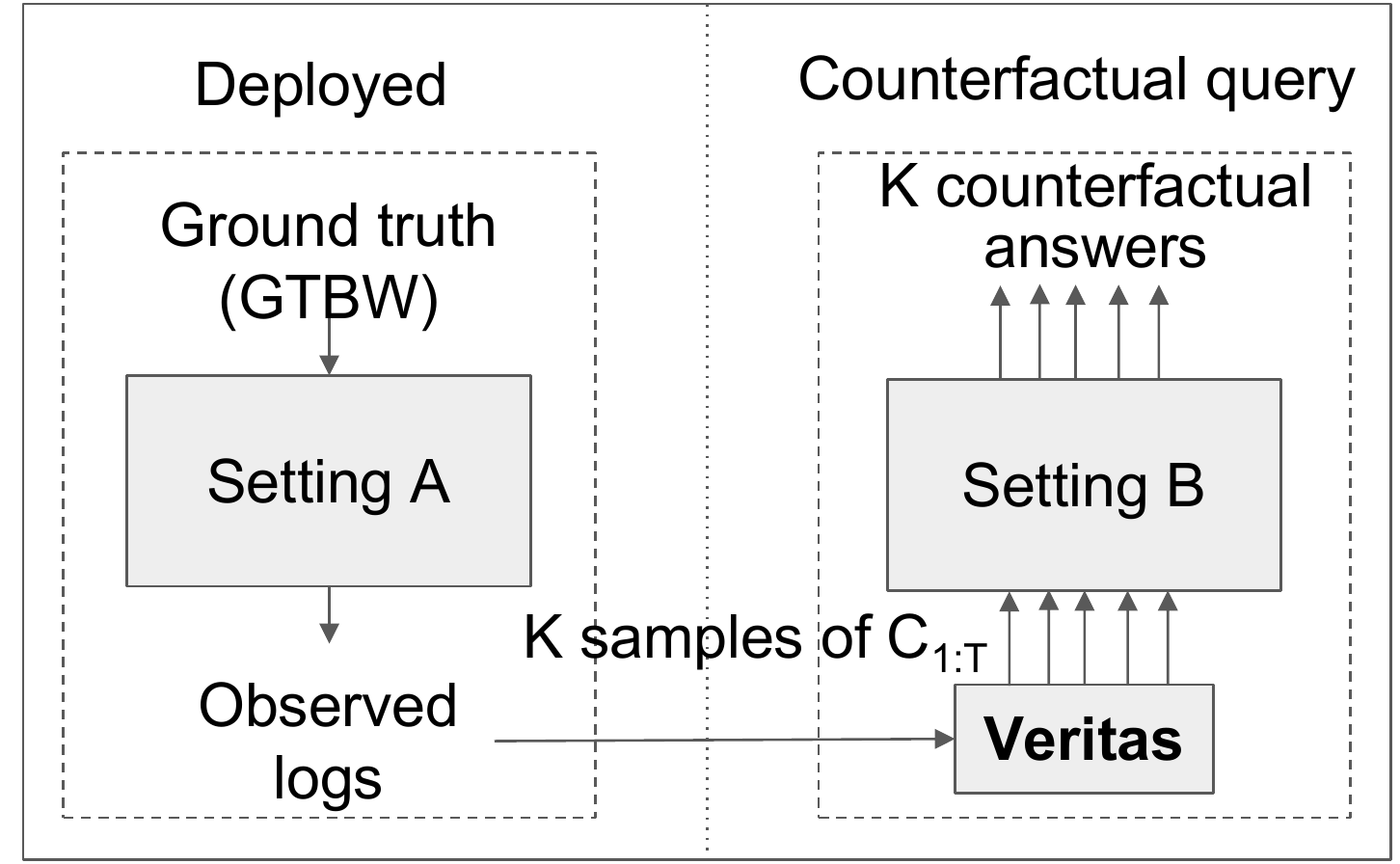}
    \caption{Using Veritas for counterfactual queries.}
    \label{fig:using-veritas}
\end{figure}

\subsection{How \System{} answers causal queries}
\label{sec:design:causalQueries}
\Cref{fig:using-veritas} shows how \System{} may be used to answer counterfactual and interventional queries. The system deployed in the wild (Setting A) produces logs which for each chunk includes (i) size; (ii) start time of download; (iii) end time of download; and (iv) TCP state including cwnd, ssthresh, and rto~\cite{tcpinfo}. 
\System{} performs the abduction step, which allows for 
the sampling $K$ likely \capacity sequences $C_{1:T}$ through \Cref{eq:C1T}. The video session is emulated in a new Setting B corresponding to the counterfactual query (e.g., Setting B may correspond to a different algorithm, or buffer size) by replaying each of these sample traces $C_{1:T}$. \System{} then provides $K$ outcomes for the counterfactual query rather than just a single one, capturing the uncertainty inherent in the abduction step given the observed data.
While the above description pertains to counterfactual queries, 
\System{} can also be used for interventional queries as we describe in \S\ref{sec:eval:interventional}.




\section{Evaluation}
\label{sec:evaluation-setup}

In this section, we evaluate \System{} with respect to how effectively it can respond to causal queries ({\em what-if} questions) when given traces collected from a video streaming system. We start by evaluating \System{}'s ability to tackle counterfactual queries (\S\ref{sec:counterfactual}) and then evaluate its effectiveness in handling interventional
queries (\S\ref{sec:eval:interventional}).

\subsection{Evaluation with counterfactuals}
\label{sec:counterfactual}
Given traces collected from a video streaming system with a particular set of design decisions, we evaluate the effectiveness of \System{} when answering {\em what-if} questions related to the performance of the system on \textit{the same set of traces} if one could go back in the past and use a new set of decisions. The decisions that we consider include (i) changing the set of video qualities that the streaming algorithm may choose from; (ii) changing the buffer size available to the video player; and (iii) changing the underlying ABR algorithm itself.

\textbf{Schemes compared.}
We compare \System{}'s ability to handle counterfactuals with two approaches: 

\noindent
$\bullet$ \textbf{Ground Truth (\GT{}):} This refers to the ground truth bandwidth, defined in \S\ref{sec:causalDAG}. When answering {\em what-if} questions, results using this technique serve as the ideal benchmark, that \System{} and other approaches must seek to achieve.  

\noindent
$\bullet$ \textbf{\Baseline{}:} This scheme directly uses the observed throughput of each chunk, and assumes this throughput value holds from the start time of the chunk download to the end time of download. During off periods when no estimate is available, linear interpolation of the throughput observed by the previous and next chunks is used. The scheme is commonly used in most video streaming evaluations today. It is expected to be a more accurate representation when the observed throughput is close to \capacity{}, but underestimates otherwise, and may be inaccurate during off-periods.  


\textbf{Evaluation setup.}
When evaluating counterfactual questions, we use the evaluation setup similar to Figure~\ref{fig:using-veritas}. First, we run a video streaming session in Setting A emulating a ground truth network bandwidth (\Capacity{}) trace, which results in a set of logs (as discussed in \S\ref{sec:design:causalQueries}). 
Next, we run the video streaming session in Setting B emulating traces approximating \Capacity{} inferred by \System{} and \Baseline{}, as well as the original \Capacity{} trace. For \System{}, we sample multiple traces (5 by default), and summarize a range of outcomes. We report the performance predicted in Setting B with each of the approaches.



\textbf{Evaluation metrics.} We compare results predicted by each of \System{}, \Baseline{}, and \GT{} with respect to the actual {\em what-if} scenario. Our counterfactual questions pertain to impact of change of setting on the quality of a video session. Hence, we use standard metrics such as video quality (measured by SSIM) and rebuffering ratios. 

\label{emulation-setup}
\textbf{Setup details.} We use the evaluation setup provided by Fugu~\cite{yan_learning_2020} to run our emulation experiments with different ABR algorithms. We emulate FCC throughput traces ~\cite{fcc} using Mahimahi ~\cite{mahimahi} to play a 10 minute pre-recorded video clip with bitrate ranging from 0.1 Mbps to 4 Mbps. We use the SSIM index ~\cite{ssim} as a measure of video quality. The average SSIM index of lowest quality and highest quality are 0.908 and 0.986 respectively.  The clients are launched inside a mahimahi shell with a 80 ms end to end delay and downlink GTBW limited by FCC traces. The GTBW of FCC traces varies from 3 Mbps to 8 Mbps. 
In our evaluation, we use MPC ~\cite{mpc} as default ABR algorithm with a buffer size of 5 seconds. Veritas uses the EHMM described in \S\ref{sec:abduction} with  \capacity transition interval size $\delta$=5s and minimum \capacity discrepancy $\epsilon$=0.5 Mbps, variance $\sigma$=0.5, a tridiagonal transition matrix $A$ and a uniform initial distribution $u$ through all capacity states. The tridiagonal transition matrix prioritizes \capacity states to be stable, but it allows variation over time. Veritas uses the throughput estimator described in \S\ref{sec:abduction}
\sgr{revisit this para: e.g., should we get into why this bandwidth range, implications of tridiagonal etc.? not sure.. could cut both ways}

\subsection{Inference with \System{}: Example}

\begin{figure*}[!htb]
\centering
\subcaphangtrue
\subfigcapmargin = .09cm
\subfigure[GTBW and Baseline.]{
\includegraphics[width=0.48\textwidth]{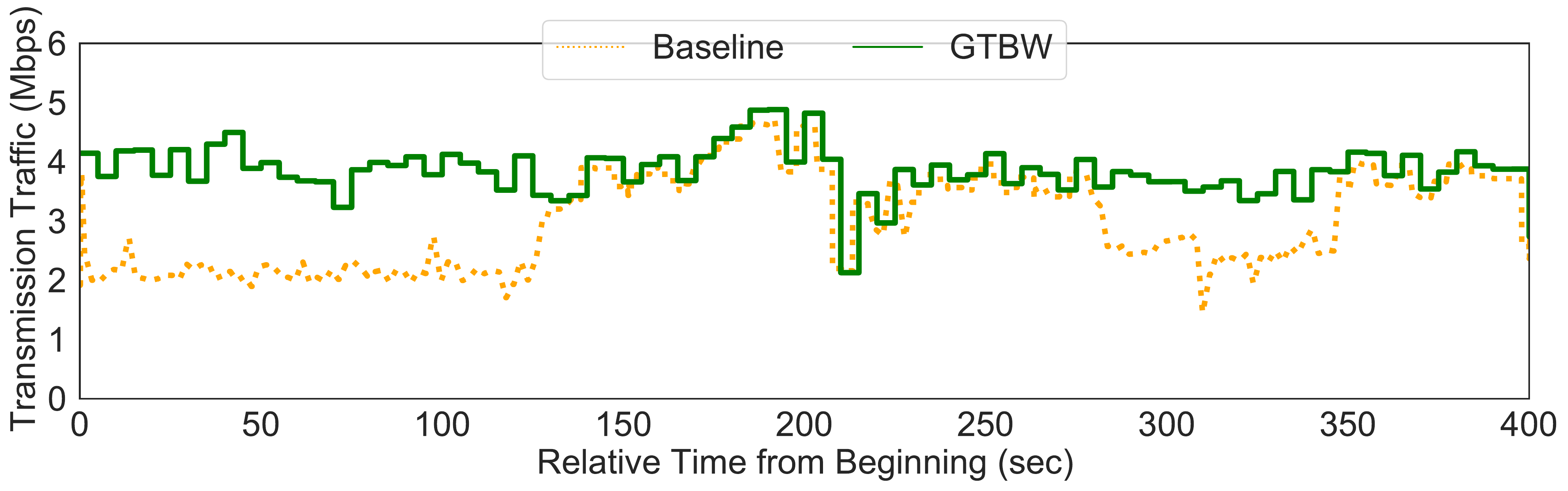}
\label{fig:what-if-step-1}
}
\subfigure[GTBW and Veritas samples.]{
\includegraphics[width=0.48\textwidth]{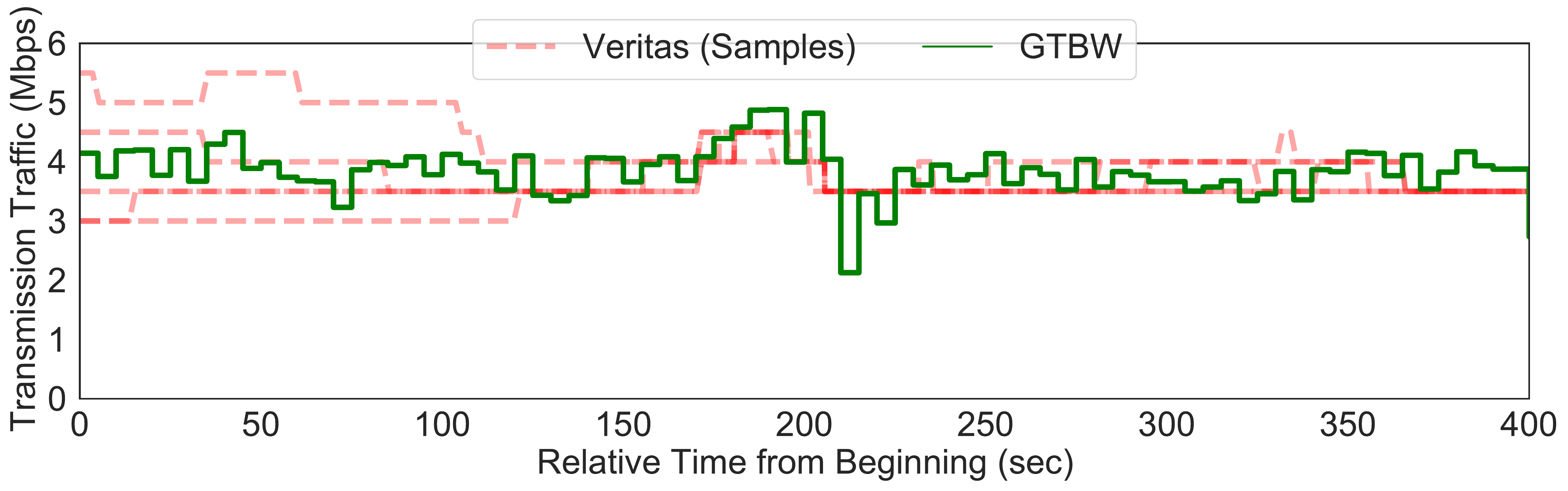}
\label{fig:what-if-step-2}
}
\caption{Comparing Baseline, \Capacity{} and \System{} samples for an example trace.}
\label{fig:example-what-if}
\end{figure*}
   


We start by illustrating \System{}'s ability to more accurately infer the \Capacity{} time series compared to \Baseline{} with an example. Figure~\ref{fig:what-if-step-1} illustrates the \Capacity{} seen during the session for an example trace, as well as the performance observed by \Baseline{}. There are significant periods (e.g. upto 120 seconds, and between 270 and 350 seconds)
where \Baseline{} is conservative in its estimation of \Capacity{}. This is because in these periods the deployed ABR algorithm selects smaller chunk sizes (either lower qualities, or lower-sized chunks of higher quality given variable bit rate video). Consequently, observed throughput in these periods is significantly lower than \Capacity{}. 

Figure~\ref{fig:what-if-step-2} illustrates the time series reconstructed by \System{} for the same \Capacity{} trace. \System{} does not provide one single trace, but allows sampling of multiple candidate traces, with more probable traces having a higher likelihood of being sampled. The figure shows five sample traces from \System{}. We make several observations. First, all these samples are closer to \Capacity{} than \Baseline{}. Second, in regions where \Baseline{} is close
to \Capacity{} (e.g., between 120 and 270 seconds), all samples from \System{} 
are also close to \Capacity{}. This is because in these regions, the chunk sizes selected by the deployed algorithm exceed the bandwidth delay product (BDP) of the network. Here, the observed throughput is closer to \Capacity{}, and \System{} is relatively more certain. Third, in regions where \Baseline{} is conservative, all \System{} samples are significantly less conservative. However, in some of these regions (e.g., 0 to 120 seconds), \System{} exhibits more uncertainty. This occurs because if smaller chunk sizes are chosen by the deployed algorithm, a range of different \Capacity{} values may have resulted in the same throughput observations. This is intrinsic, reflecting the uncertainty inherent in the available data. Note that \System{}'s use of HMMs allows it to pick more probable samples based on transition probabilities (i.e., since it infers \Capacity{} in some regions with higher certainty, the transition probabilities constrain the range of \Capacity{} possibilities in the less certain regions).



\begin{figure}[h]
\centering
\subcaphangtrue
\subfigure[SSIM]{
\includegraphics[width=0.22\textwidth]{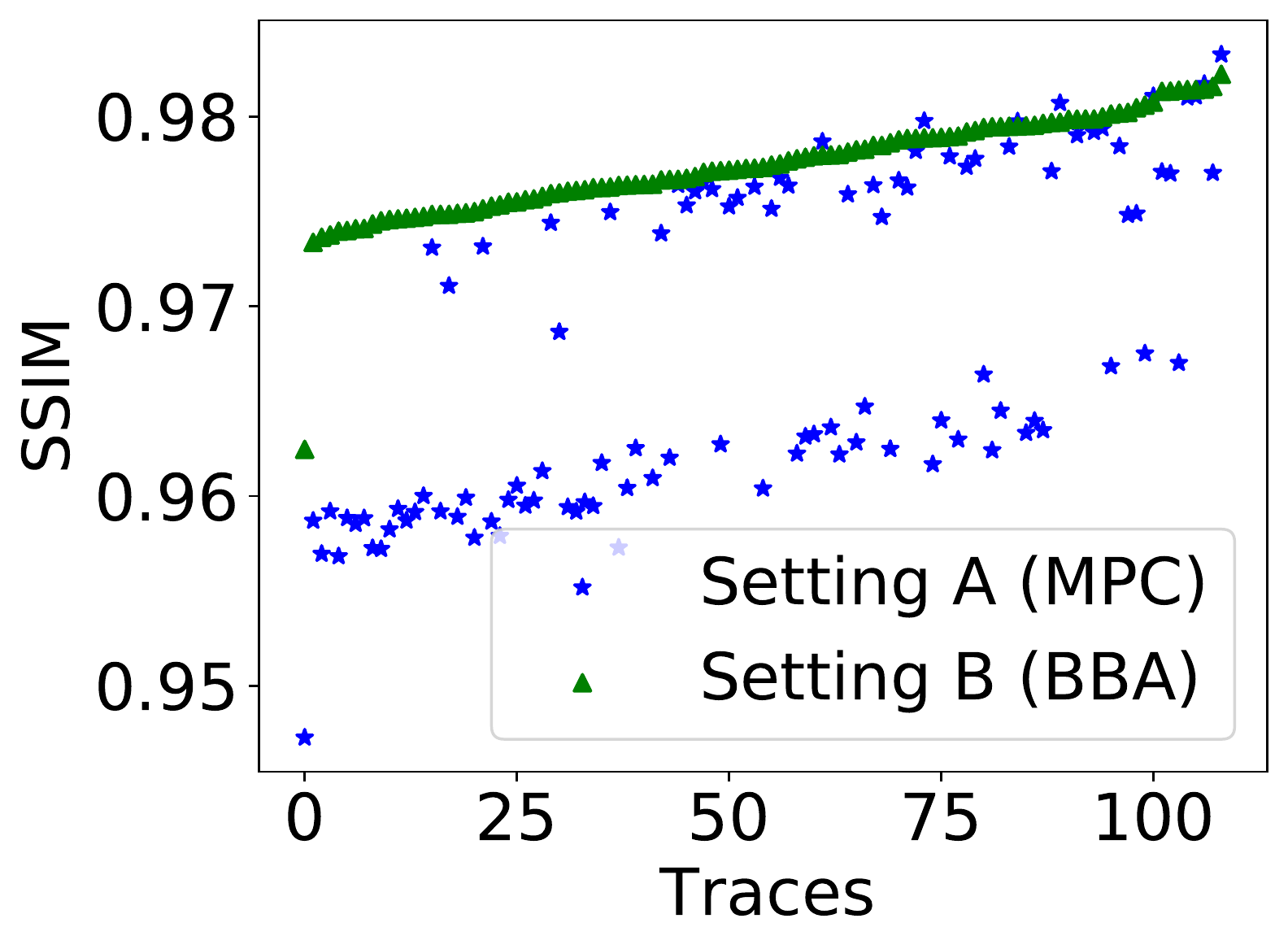}
\label{fig:ssim-A-B}
}
\subcaphangtrue
\subfigure[Rebuffering]{
\includegraphics[width=0.22\textwidth]{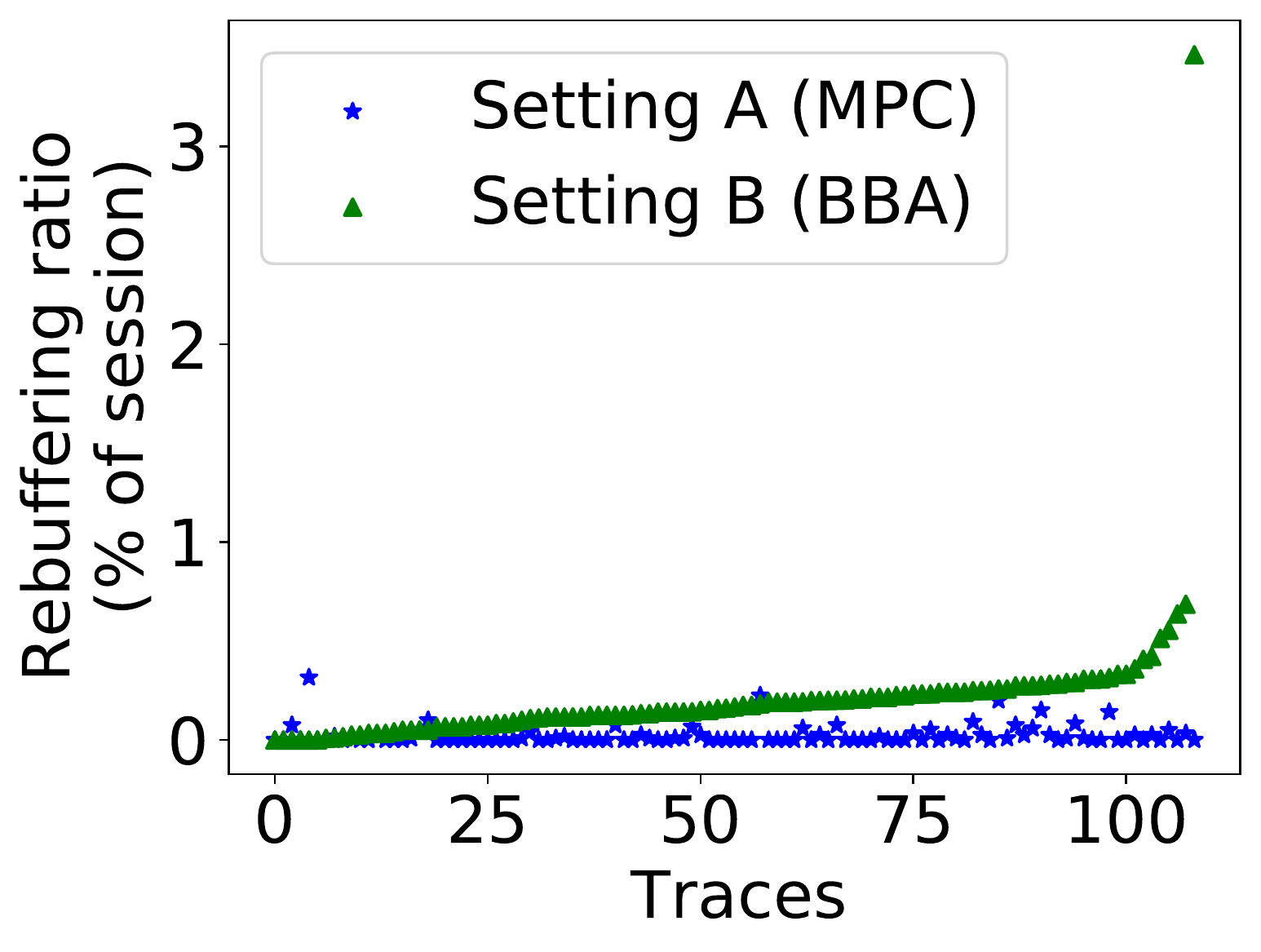}
\label{fig:rebuf-A-B}
}
\caption{True impact of changing ABR algorithm from MPC to BBA.}
\label{fig:deployment-gt}
\end{figure}

\begin{figure}[h]
\centering
\subcaphangtrue
\subfigure[SSIM]{
\includegraphics[width=0.22\textwidth]{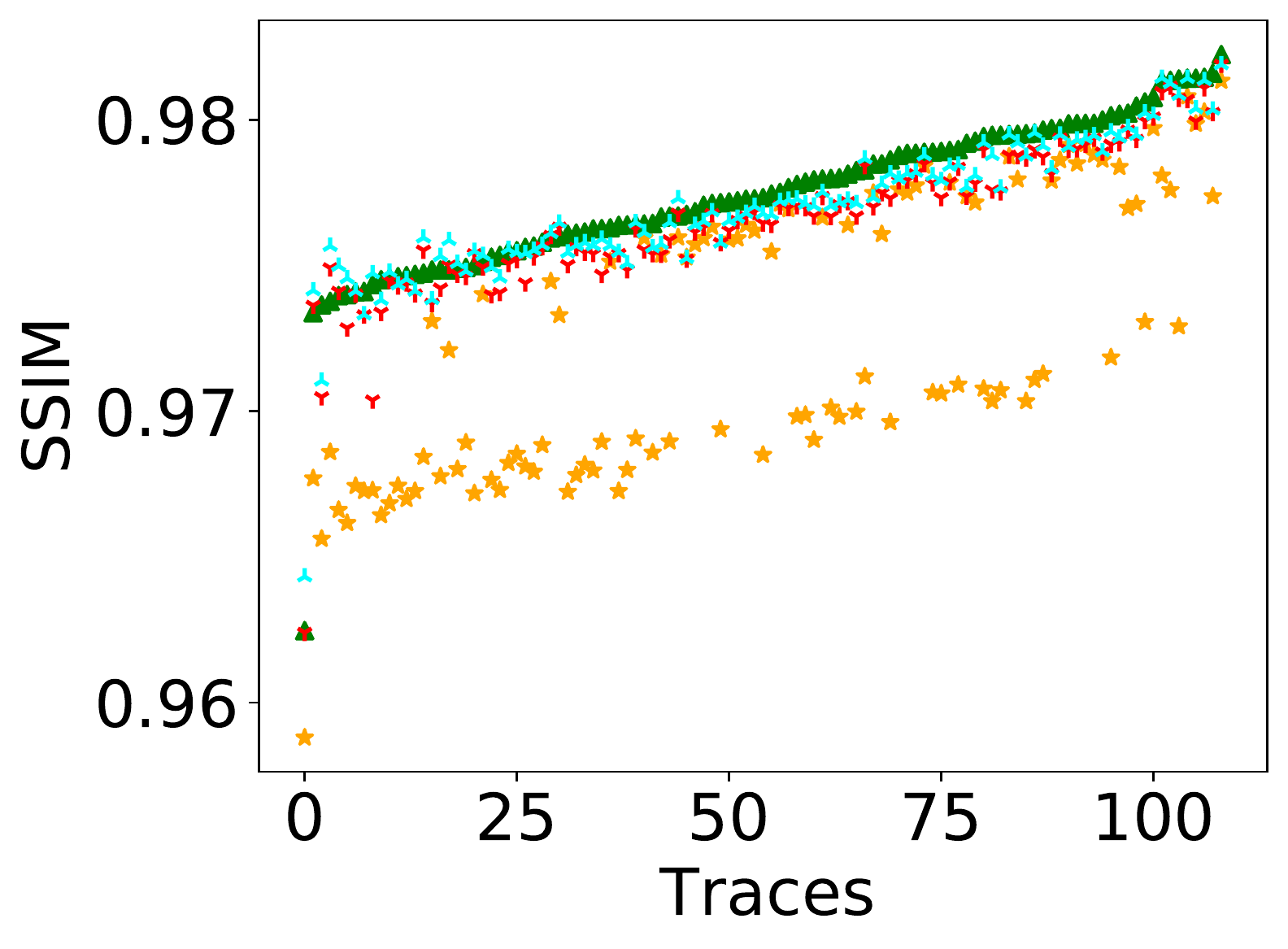}
\label{fig:change-abr-ssim}
}
\subfigure[Rebuffering]{
\includegraphics[width=0.22\textwidth]{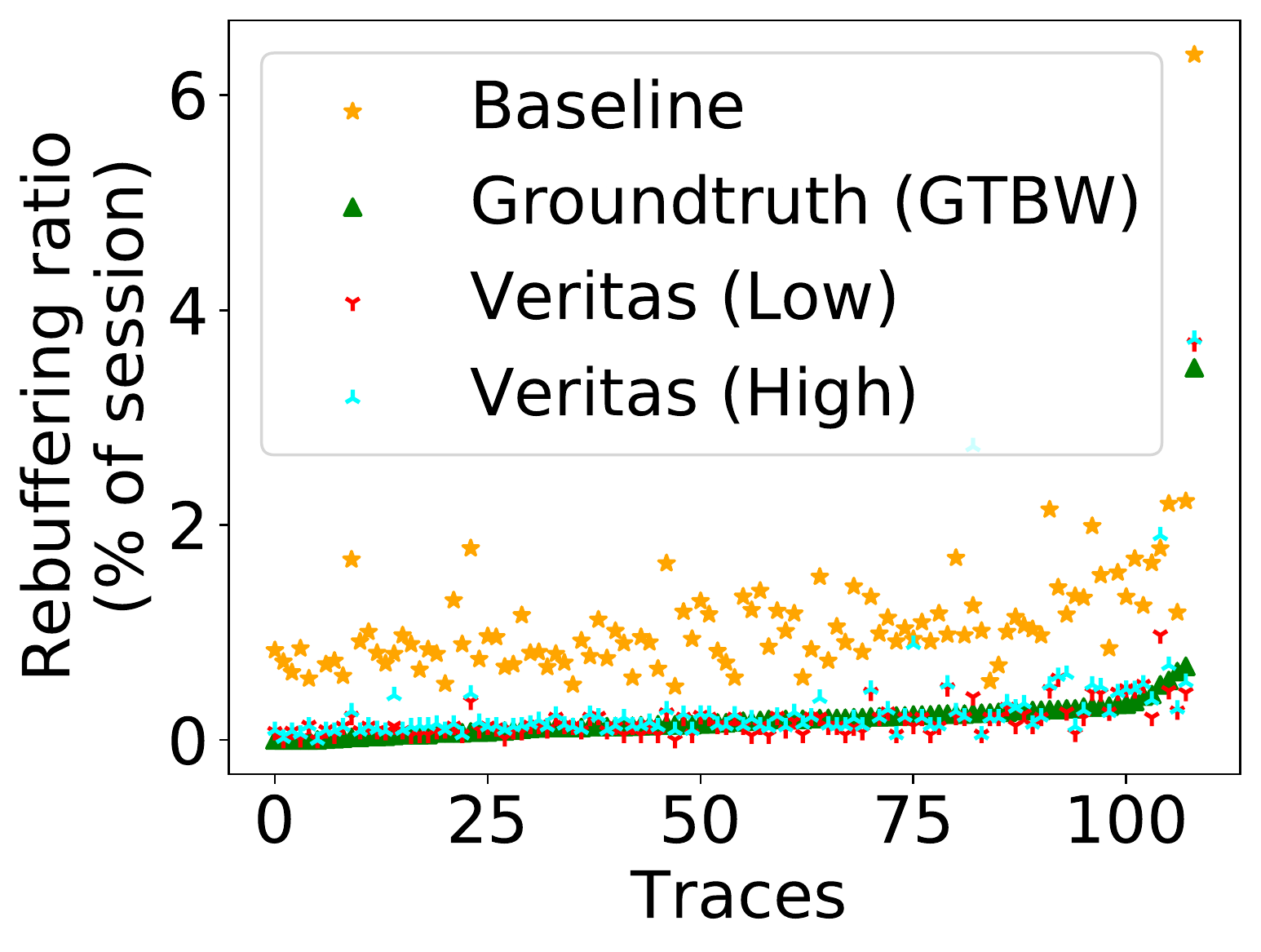}
\label{fig:change-abr-rebuf}
}
\caption{Predicted performance if ABR was changed from MPC to BBA.}
\label{fig:abr-metrics}
\end{figure}

\begin{figure}[h]
\subcaphangtrue
\subfigure[SSIM]{
\includegraphics[width=0.22\textwidth]{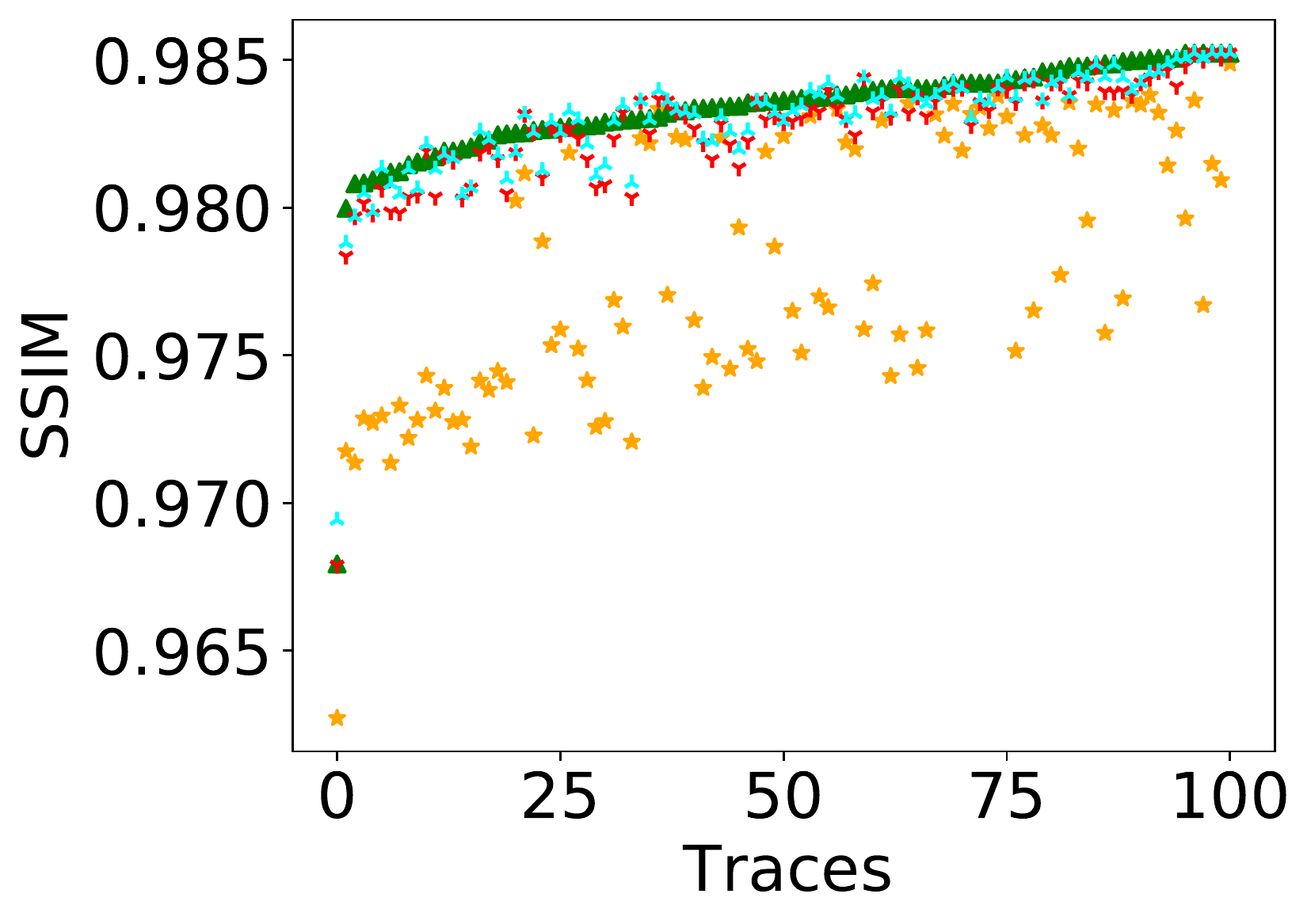}
\label{fig:change-buffer-ssim}
}
\subfigure[Rebuffering]{
\includegraphics[width=0.22\textwidth]{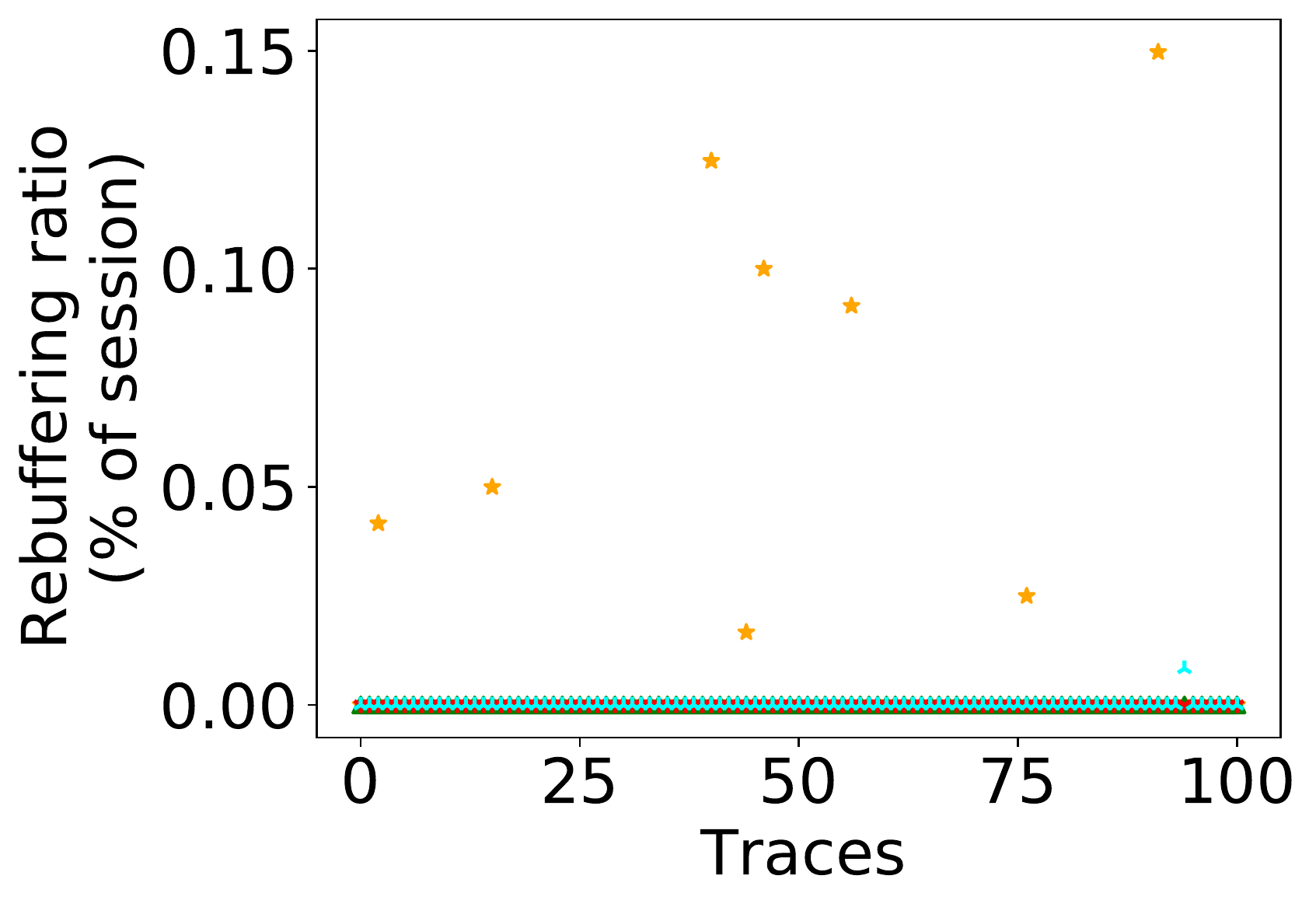}
\label{fig:change-buffer-rebuf}
}
\label{fig:buffer-change}
\caption{Predicted performance if buffer size was increased to 30s.}

\end{figure}

\begin{figure}[h]
\subfigure[SSIM]{
\includegraphics[width=0.22\textwidth]{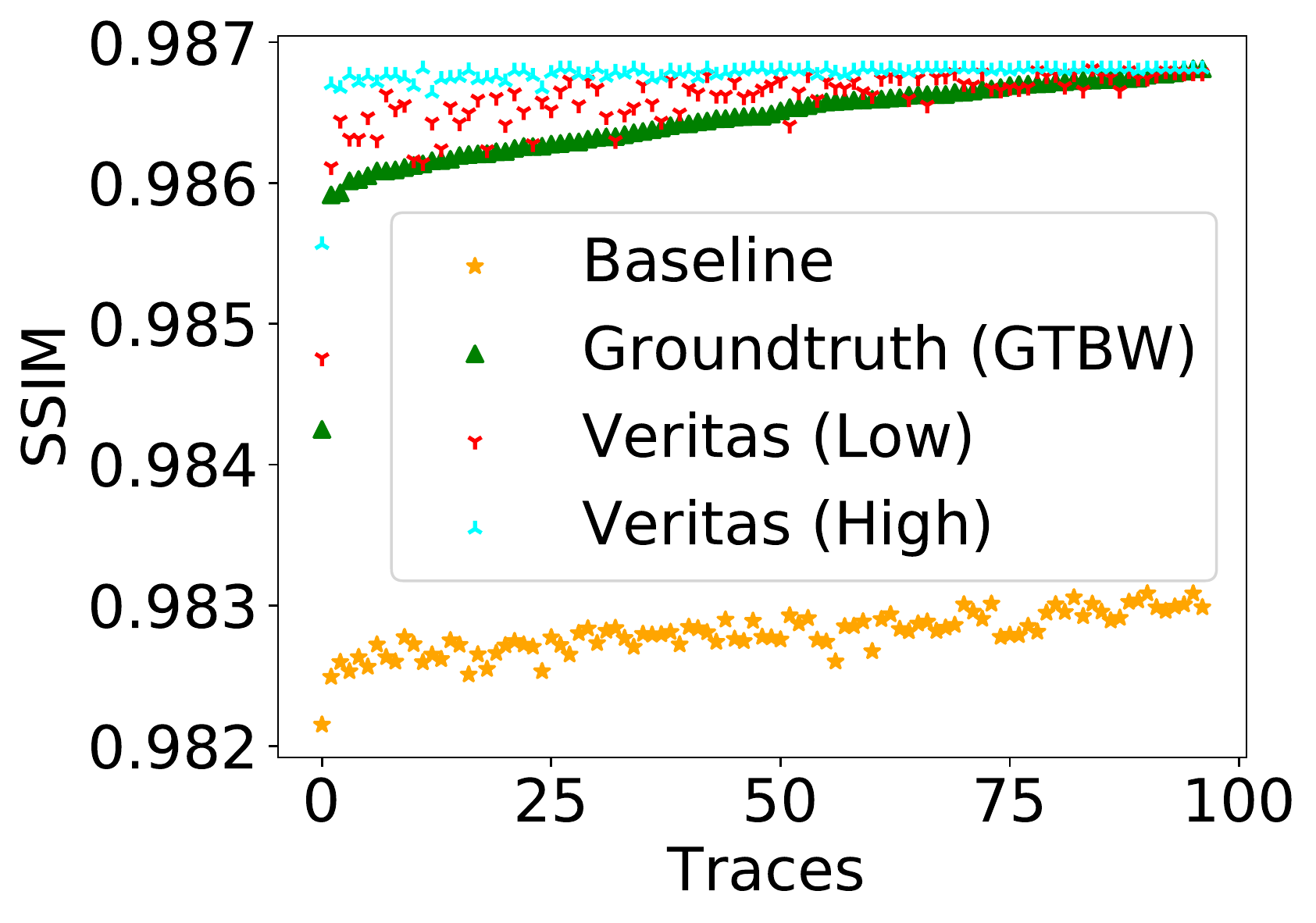}
\label{fig:change-quality-ssim}
}
\subfigure[Rebuffering]{
\includegraphics[width=0.22\textwidth]{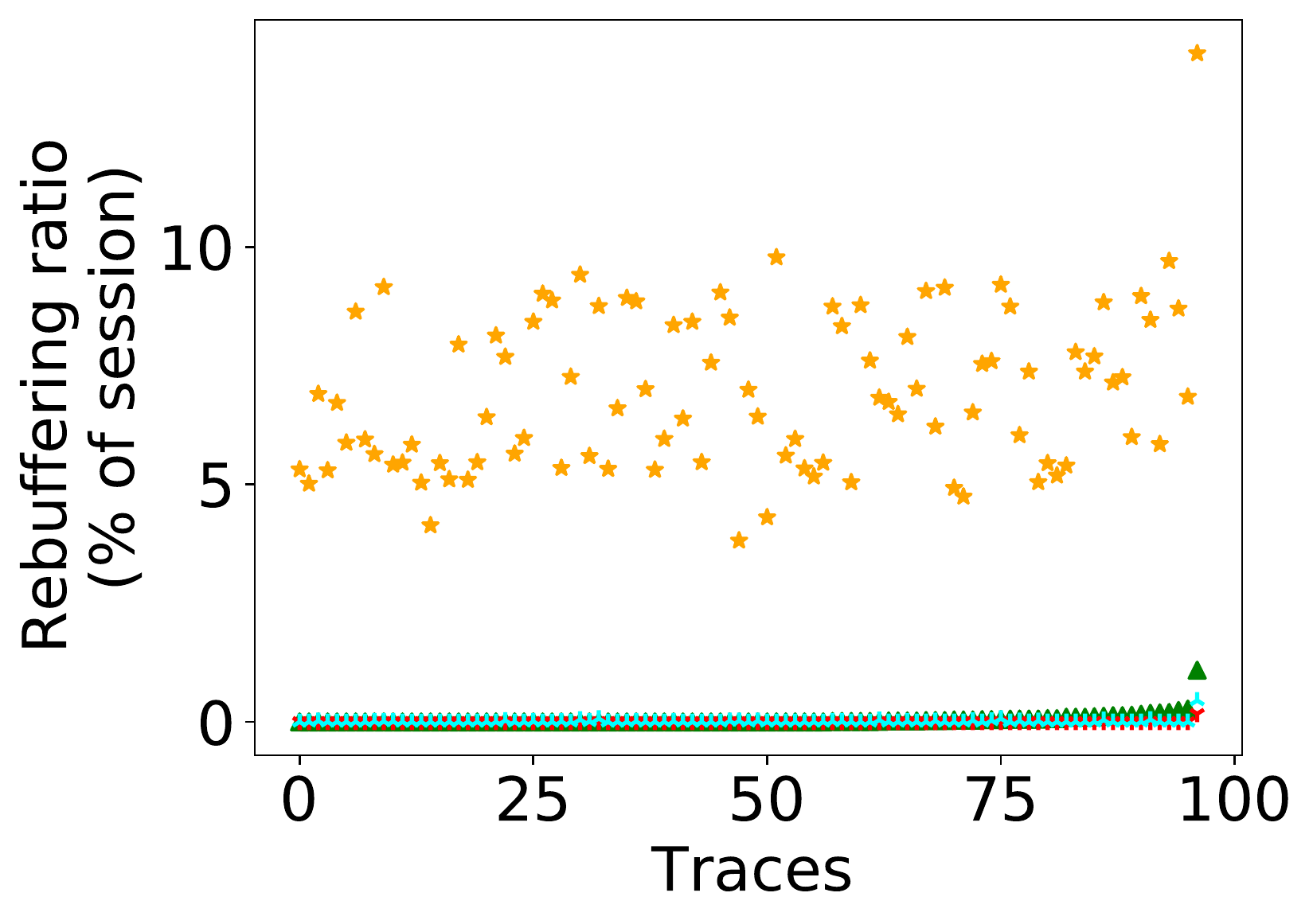}
\label{fig:change-quality-rebuf}
}
\caption{Predicted performance if higher video qualities were used.}
\label{fig:quality-metrics}

\end{figure}




\subsection{Results: \System{} with counterfactuals}
We next evaluate \System{}'s ability to answer three example counterfactual questions.

\noindent
\textbf{Change of ABR algorithm.}
Consider that the video streaming application has been deployed with a given ABR. We ask the counterfactual {\em what would have happened if an alternate ABR algorithm were instead used}. We study this question in the context of moving from the MPC algorithm~\cite{mpc} to the BBA algorithm~\cite{tysigcomm2014}. \Cref{fig:ssim-A-B,fig:rebuf-A-B} show the SSIM and rebuffering ratio achieved by MPC and BBA when using the same set of \Capacity{} traces. In each graph, each point on the X-Axis corresponds to a \Capacity{} trace, and each graph plots the SSIM (or rebuffering ratio) for that trace with the two algorithms. Notice that BBA is more aggressive with larger SSIM values and higher rebuffering. \sgr{Why} \sgr{More informative legends/captions.}



We next evaluate the ability of \Baseline{} and \System{} to predict the impact on video performance if (i) logs from a deployment of the MPC algorithm were provided; and (ii) the BBA algorithm were used instead.
For each video session, we infer five
sample \Capacity{} time series using \System{},
and emulate each video session under each of the five \System{} samples. Each sample provides a prediction of SSIM and rebuffering with \System{}.
We consider the second lowest and second largest prediction for each metric across the samples,
which we refer to as \System{} (Low) and \System{} (High) respectively. This provides a range of predictions with \System{} for each trace.

\Cref{fig:change-abr-ssim,fig:change-abr-rebuf} present the SSIM and rebuffering ratio predicted by \Baseline{} and \System{}. The true impact of the change (\Capacity{}) is also shown for comparison.
The graph shows that \Baseline{} predicts a noticeably lower SSIM than \Capacity{}\footnote{Average bit rate for the median trace reduces from the true  value of 3.5 Mbps to 3.1 Mbps for baseline. See Appendix which presents the impact on average bit rate for all counterfactual queries.}, and a significantly higher rebuffering ratio. This is because \Baseline{} underestimates \Capacity{} as we have seen.  In contrast, the range of estimates from \System{} is close to \Capacity{} across the traces and fairly tight indicating \System{} is confident in assessing the impact of this change. 

In the Appendix, we have also evaluated the impact of changing from MPC to the BOLA algorithm, which shows similar results: \System{} does a good job of predicting the impact of the change, but \Baseline{} does not. \sgr{Add BOLA graphs in Appendix. Fix references}


\noindent
\textbf{Change of buffer size.}
Consider that the video streaming application has been deployed with an ABR and a buffer size. The designer then asks: {\em what would have been the performance if a different buffer size had been used?} Intuitively, increasing the buffer size should improve video quality and lower rebuffering, but lower the liveness for the application.
We deploy the MPC algorithm with a buffer size of 5 seconds (Setting A), and using the logs so obtained, evaluate the impact predicted by different schemes if the buffer size were increased to 30 seconds (Setting B). \Cref{fig:change-buffer-ssim,fig:change-buffer-rebuf} show the results.
\System{} accurately predicts SSIM and rebuffering ratio (close to \Capacity{}), with the range of estimates for each trace being relatively tight. \Baseline{} underestimates SSIM for most traces, and slightly over-estimates rebuffering ratios for some traces. 


\noindent
\textbf{Change of qualities.}
Consider that the video streaming application has been deployed with a given set of video qualities.
We next consider the counterfactual: {\em what would be the impact if a higher set of qualities were used instead?} \sgr{Talk about quality used or not?}
%
\Cref{fig:quality-metrics} shows that \System{} achieves SSIM and rebuffering close to \Capacity{}. However, \Baseline{} underestimates SSIM and significantly overestimates rebuffering (the estimates
of rebuffering ratio with \Baseline{} are in the 5-10\% range across traces, while the estimates are close to
0 with \Capacity{} and \System{} for most traces). Note that for this case study, \System{} tends to slightly over-estimate SSIM relative to \Capacity{}. This is because in most traces in the deployment, the downloaded chunk sizes were under the bandwidth delay product, leading to a wide range of possible \Capacity{} time series consistent with the observed throughput values. Such variance is inherent to the information in the data.
\System{} can provide a range of outcomes in general, and obtaining more samples could potentially lead to lower estimates. Overall, \System{} is effective in answering the counterfactuals and far more accurate than \Baseline{}.


\subsection{Evaluations on interventionals}
\label{sec:eval:interventional}
So far, we have evaluated \System{} on counterfactual queries that involve evaluation on a trace if one could
go back to the past and change the setting. We next evaluate \System{}'s potential for interventional queries,
which relate to the future (\S\ref{sec:background}), focusing on the ability to predict chunk download times in 
a bias-free fashion. We compare two schemes:

\noindent $\bullet$ \textbf{FuguNN:} This refers to a neural network proposed in \cite{yan_learning_2020} which predicts the download time of chunks based on the sizes and download times of prior chunks. While the approach is effective at predicting chunk download times for 
sizes selected by the deployed ABR, it suffers from a bias when predicting download times for alternate chunk sizes different than what
the deployed ABR may have selected as shown in \S\ref{sec:causalQueries}.

\noindent $\bullet$ \textbf{\System{}}: Using only the chunks downloaded upto a particular point in the session, we use \System{} to infer \Capacity{} time series for the past. We consider a single sample from \System{} corresponding to the most likely one.
We then use the 
transition matrix to get the expected value of \capacity{} for the next chunk. 

    

      
\begin{figure}[h]
    \centering
    \includegraphics[width=5cm]{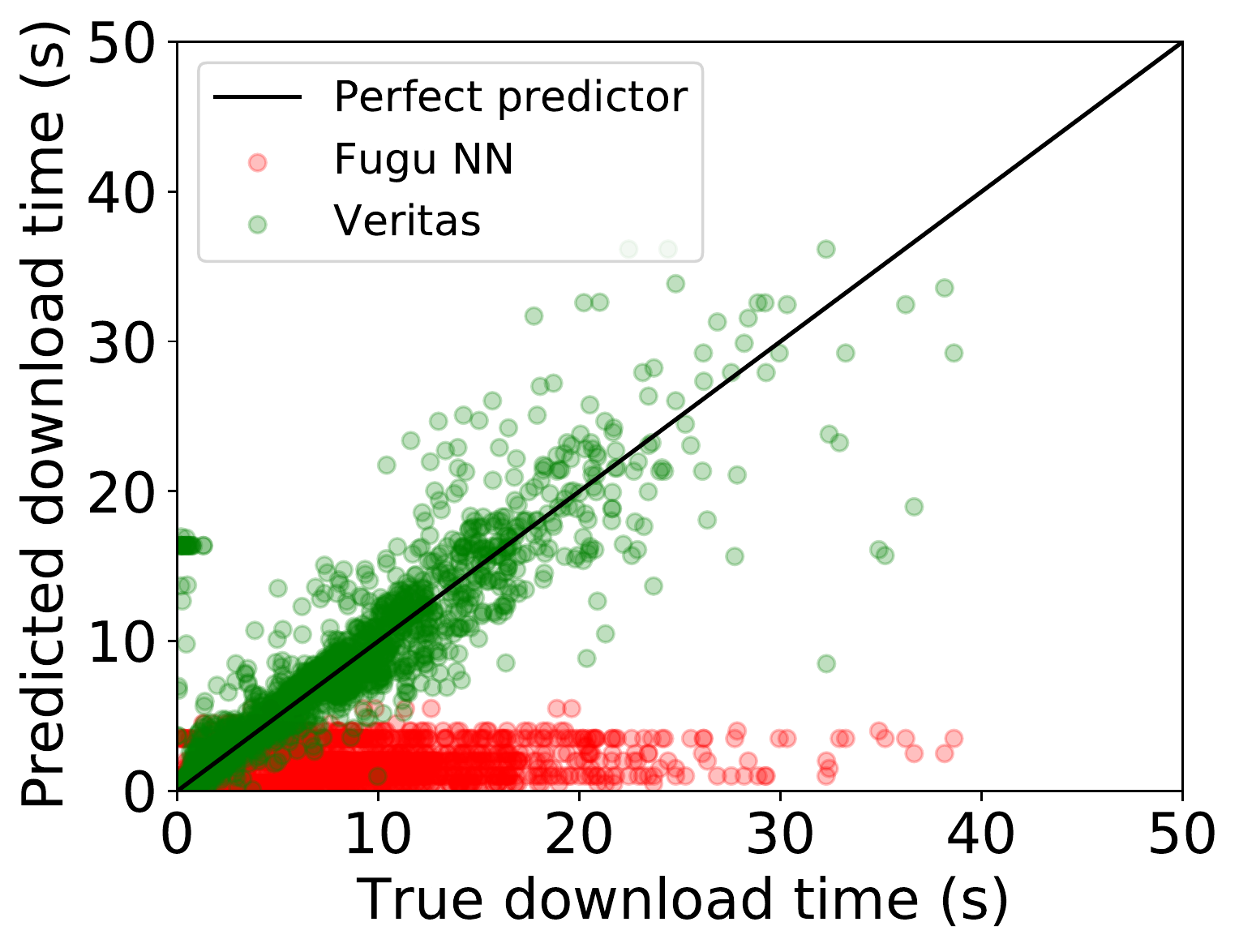}
    \caption{Comparing FuguNN and HMM for download time prediction in an interventional context.}
    \label{fig:exp-fugu-bias-result}
\end{figure}

We train FuguNN using traces obtained by running the MPC algorithm on 100 FCC traces sampled uniformly from all traces with average \Capacity{} values ranging from
0.5 to 10 Mbps.  We then create a separate set of 30 traces drawn from the same range of \Capacity{},
but where bit rates are selected randomly rather than use an ABR algorithm to serve as a test set. The purpose is to evaluate
FuguNN and \System{} on their ability to predict chunk download times for arbitrary chunk sequences, not specific to what one may
encounter in the deployed algorithm. Each test trace gives us ground truth information for the size and download time for several
chunk sequences.
We then run FuguNN and \System{} offline for every chunk sequence of each trace, and obtain the download time predicted by each of 
these methods. The results are shown in Figure \ref{fig:exp-fugu-bias-result}. FuguNN underestimates the download time due to 
its associational model and is unable to estimate correctly for interventional queries that involve chunk sequences different 
than what one might expect with the deployed ABR. \System{} however can effectively handle such interventional queries.
Note that as discussed in \S\ref{sec:challenge}, this is important when using using FuguNN as a predictor in a live
session since at each time step, it is used to predict the download times for \textit{all possible chunk sizes} (not just the size
the deployed algorithm would have selected).


\section{Related Work}

\noindent
$\bullet$ \textbf{Biases with video streaming.}
A very recent parallel work~\cite{alomar2022causalsim} and a preliminary workshop paper~\cite{sruthi2020netai} are motivated by similar goals as this paper. However,~\cite{sruthi2020netai}  is restricted to a square wave bandwidth process, does not model the dependence of observed throughput on chunk size, or handle the uncertainty in inference. Finally, the use of matching in~\cite{sruthi2020netai} requires bitrates to be occasionally chosen randomly.
\cite{alomar2022causalsim} has an RCT requirement in the training phase where each of $N$ sessions is assigned to one of $K$ ABR policies completely at random, and proposes counterfactual estimation as a matrix completion task. As discussed in ~\Cref{sec:RCT}, it is unclear how such approaches can evaluate actions that were outside the scope of the initial RCT experiment (e.g., what if the ABR now allowed 8K videos, or could use a different buffer size?). Moreover, deploying RCT ABR algorithms to collect traces can impact the performance of real-world users. In contrast, our work does not require RCT traces and can answer any {\em what-if} query without constraints.

Another work~\cite{bartulovic_biases_2017} has observed that smaller chunk sizes may see poorer throughput than larger ones owing to TCP slow start effects. To handle this,~\cite{bartulovic_biases_2017} compares the total reward seen by algorithm $B$ on a trace collected from an algorithm $A$ by only considering those chunks where the new algorithm picks the same bitrate as the old algorithm. The approach does not tackle {\em what-if} questions, assumes a constant bandwidth process, and does not model the causal dependence of chunk size selection by the ABR algorithm on bandwidth. 
We tackle the harder problem of inferring a latent and variable bandwidth process from observed throughput, deal with the uncertainty in such inference, and address a wide range of causal {\em what-if} queries.

\noindent
$\bullet$ \textbf{Inferring causal dependencies and {\em what-if} analysis.} 
Several works~\cite{tariq_answering_2008,kobayashi_mining_2018,mahimkar_towards_2009} infer causal dependencies using correlations but do not consider latent confounders. 
Some work~\cite{krishnan_video_2012,tariq_detecting_2009, kobayashi_mining_2018,hours_causal_2015} deals with observed confounders -- e.g., Krishnan et al.~\cite{krishnan_video_2012} explored whether video stream quality (e.g., rebuffering ratios) causally impacts user engagement metrics while acccounting for observed confounders such as user connection type (DSL vs. mobile) and location. These works only infer if a correlation is an indication of a causal relationship but do not answer {\em what-if} questions, and do not deal with latent confounders. Other works~\cite{singh_analytical_2013, jiang_webperf_2016, tariq_answering_2008} consider {\em what if} analyses for various applications, but do not address confounding variables. 

Recent work~\cite{SayerSOCC21} considers causal questions while considering implicit feedback in the context of cloud systems -- e.g., when a system waits $X$ minutes for an event to occur, there is implict feedback in terms of what would have happened if a smaller wait time were used. This approach relies on randomized experiments (from RL exploration)
and, thus, does not need to explicitly consider all possible confounders. In contrast, we are interested in scenarios where randomized experiments are limited or not available.


\section{Conclusion}
In this paper, we have made three contributions.
First, we have shown causal reasoning is complex with ABR video,
since the quality of selected video chunks is causally dependent on 
\Capacity{}, which acts as a sequence of latent and confounding variables.
Second, we present \System{}, a novel framework that tackles causal reasoning for video streaming without resorting to randomized trials.
\System{} uses an embedded Hidden Markov Model that relates 
the latent \Capacity{} time series to throughput observed by the application. A key insight behind \System{} is exploiting information  about the TCP state at the start of each chunk download to simplify
the causal inference.
%
%
Third, we show the effectiveness of \System{} in answering a wide range of counterfactual and interventional queries through emulation testbed experiments. For example, when predicting the impact of using higher video qualities, 
\System{} predicts neglible rebuffering ratios, matching ground truth.
However, \Baseline{} (which does not adjust for causal effects) predicts much higher median rebuffering ratios (6.7\%). 
With interventional queries related to chunk download times, \System{} predicts download times close to true values, while Fugu's associational approach can underestimate chunk download times 
by 5.8 seconds for 10\% of the chunks, and underestimate download times by as much as 35 seconds in the worst case.

%

%

{
\bibliographystyle{plain}
\bibliography{references/isl, references/xatu, references/causal, references/hmm, references/tools}}
\clearpage
\appendix
\section{Appendix}
\label{sec:appendix}

\subsection{Further Results}
\label{app:results}

\begin{figure}[h]
\centering
\subcaphangtrue
\subfigure[SSIM]{
\includegraphics[width=0.22\textwidth]{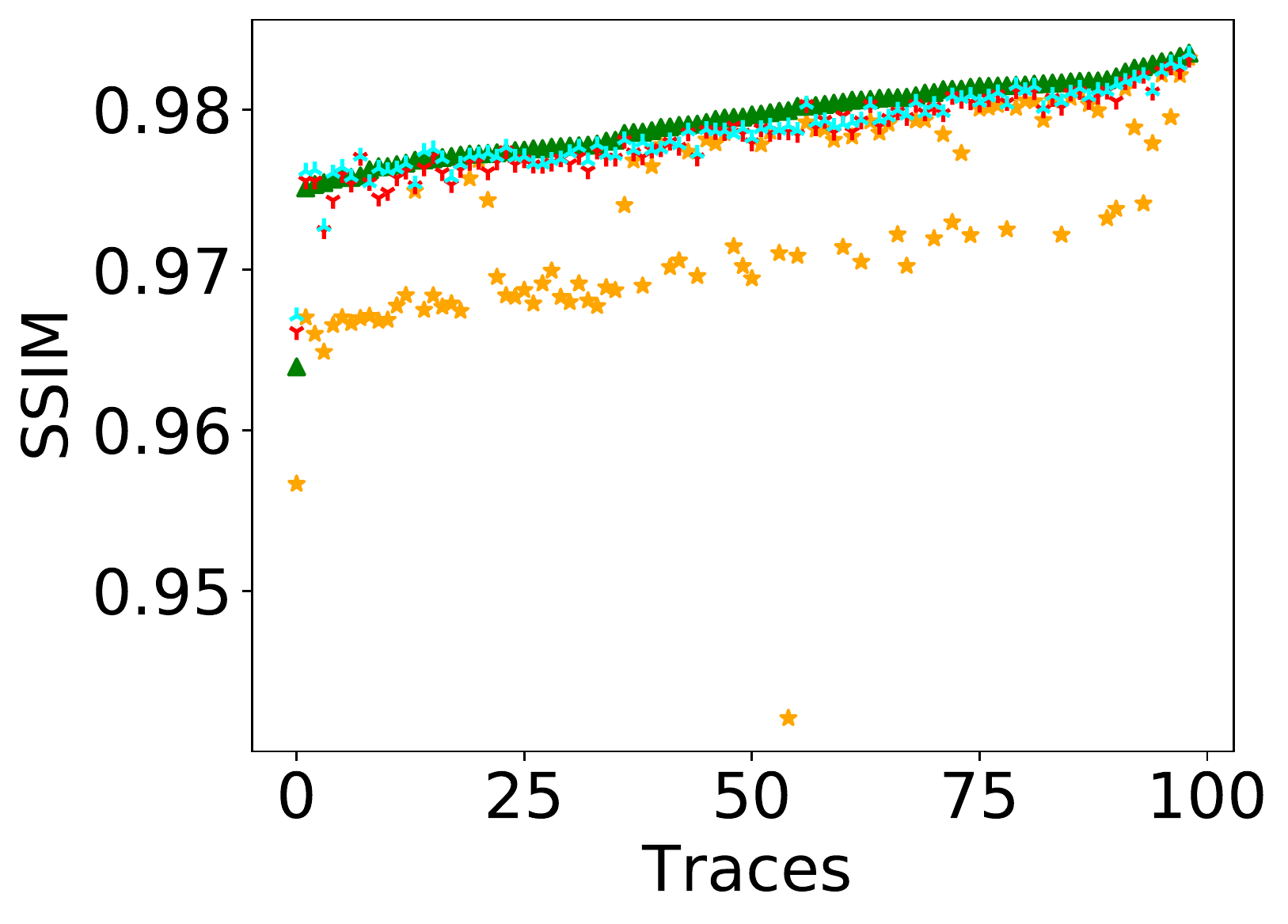}
\label{fig:appendix-bola-ssim}
}
\subfigure[Rebuffering]{
\includegraphics[width=0.22\textwidth]{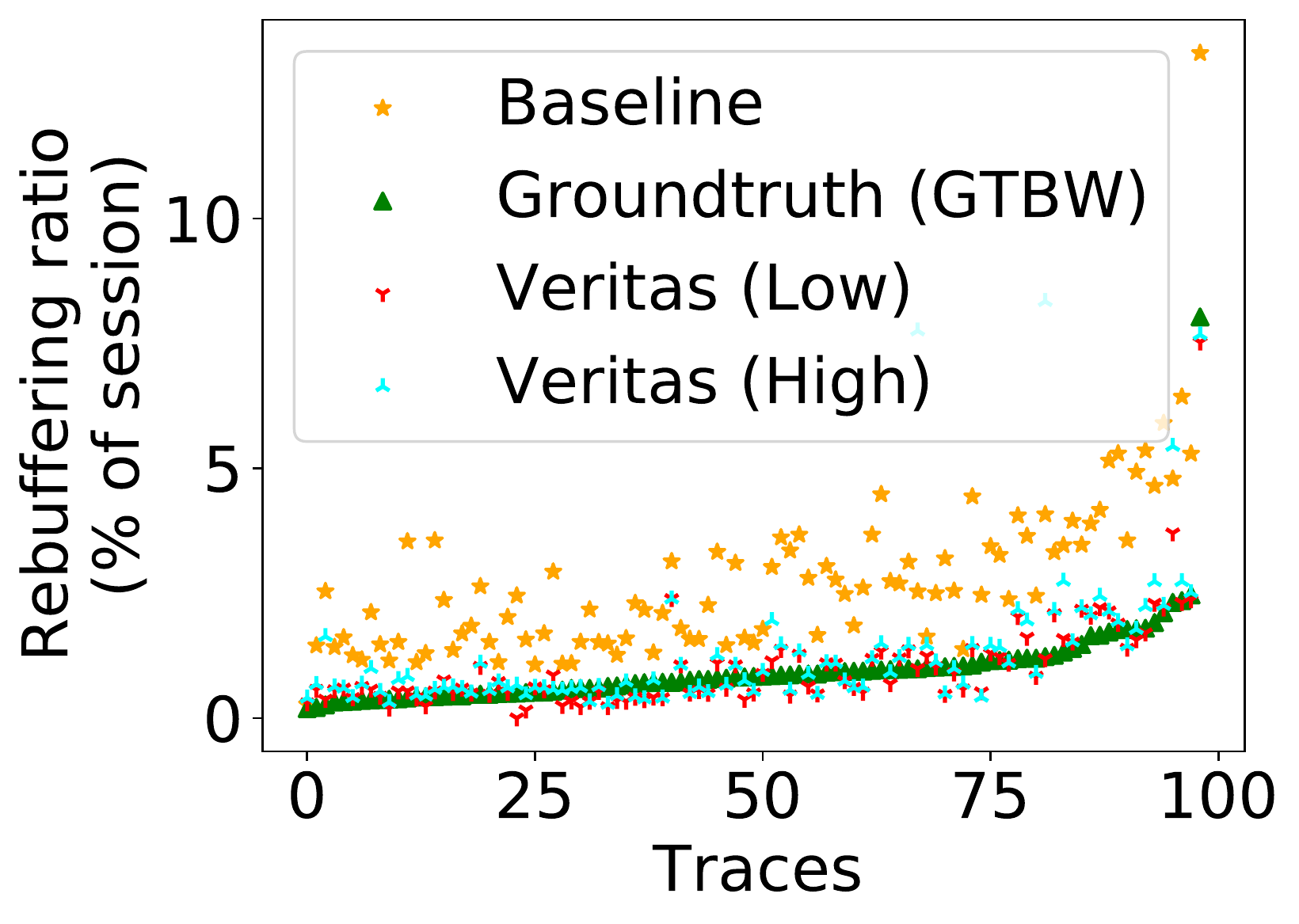}
\label{fig:appendix-bola-rebuf}
}
\label{fig:appendix-bola-metrics}
\caption{Change of ABR from MPC to Bola.}
\end{figure}

\vspace{5pt}
\noindent
\Cref{fig:appendix-bola-rebuf} and \Cref{fig:appendix-bola-ssim} show the SSIM and rebuffering ratio predicted by \System{} and Baseline when we change the algorithm from MPC and Bola~\cite{bola}. We use the Bola Basic V1 algorithm implemented in the Puffer setup for this analysis~\cite{bola-v1}. The results are similar to that of changing the ABR from MPC to BBA. Baseline underestimates the GTBW which leads to lower SSIM and higher rebuffering. \System{} does a good job of predicting the impact of the change, but Baseline does not.

\begin{figure}[h]
\centering
\subcaphangtrue
\subfigure[Avg. bitrate in Setting A and B (MPC and BBA).]{
\includegraphics[width=0.22\textwidth]{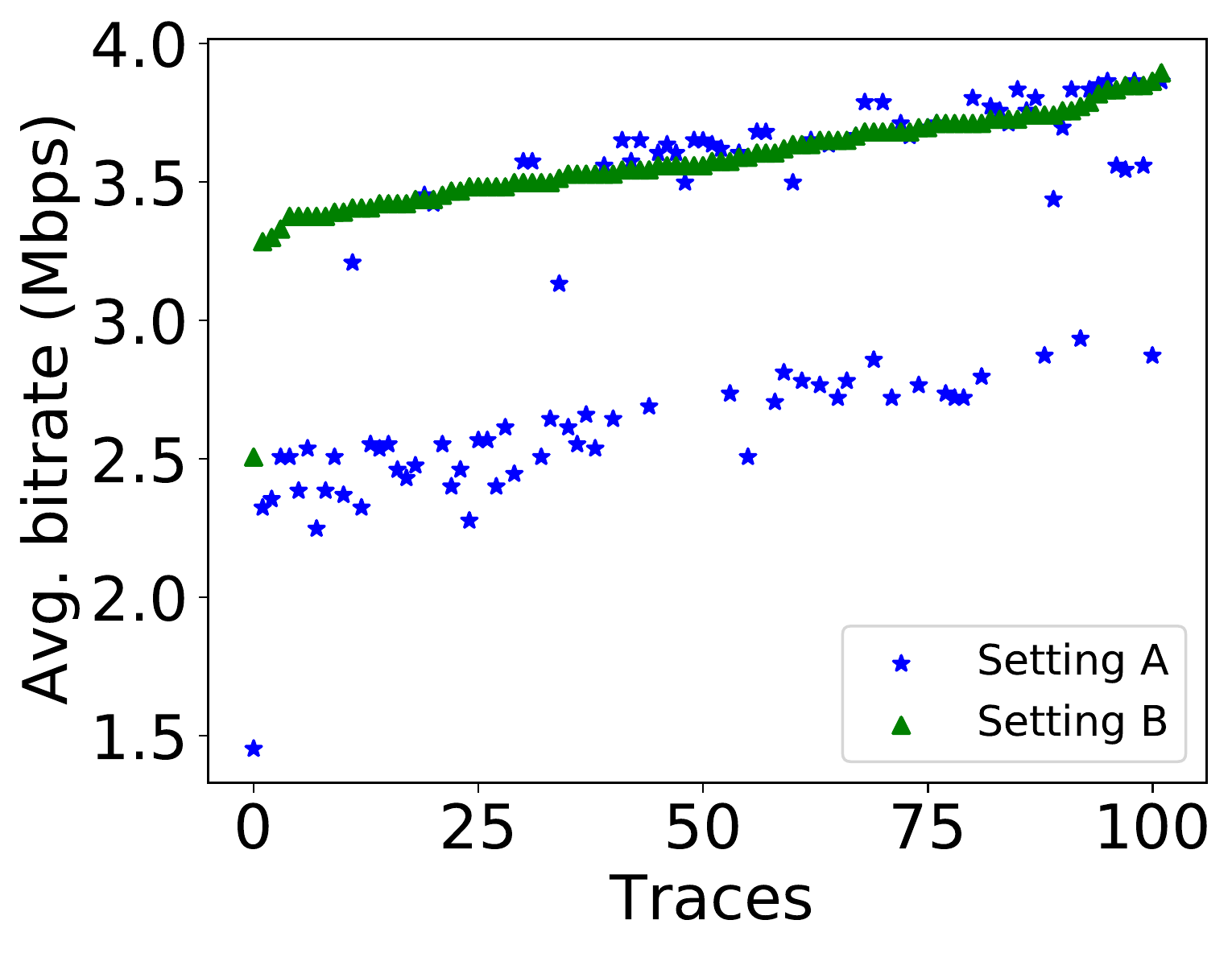}
\label{fig:appendx-bitrate-A-B}
}
\subcaphangtrue
\subfigure[Avg. bitrate for change from MPC to BBA.]{
\includegraphics[width=0.22\textwidth]
{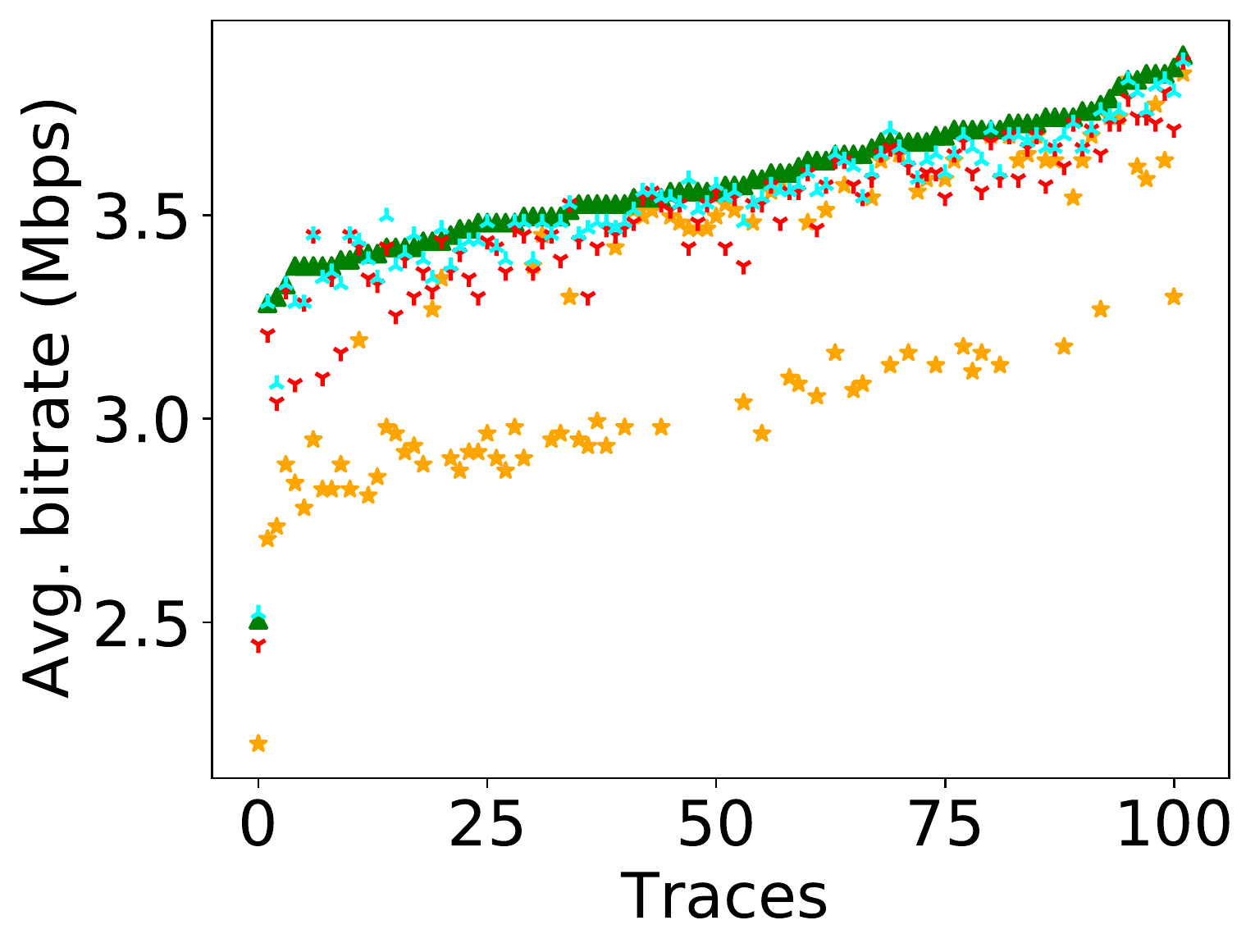}
\label{fig:appendix-bitrate-bba}
}
\subcaphangtrue
\subfigure[Avg. bitrate for change from MPC to Bola.]{
\includegraphics[width=0.22\textwidth]{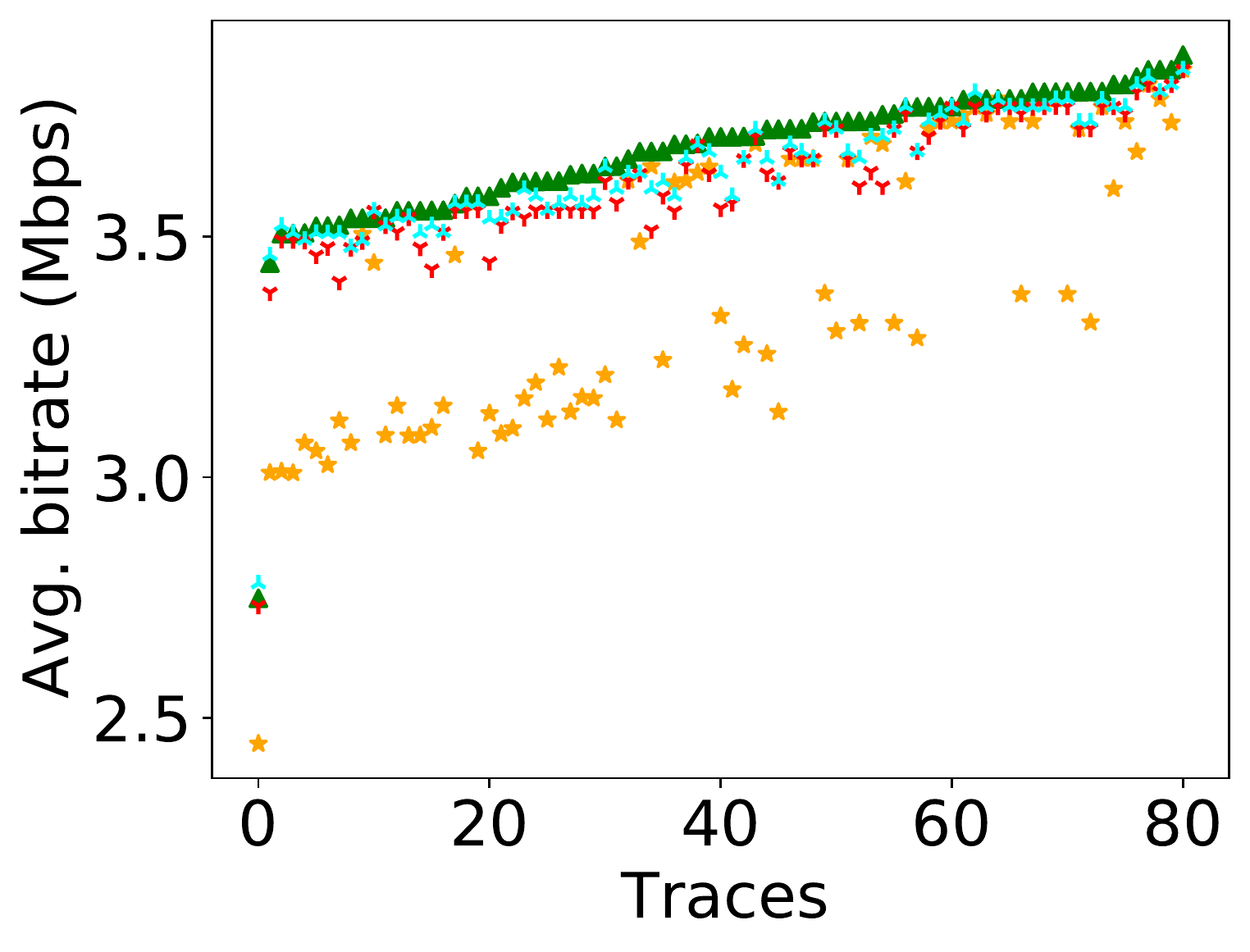}
\label{fig:appendix-bitrate-bola}
}
\subfigure[Avg. bitrate for changing buffer size.]{
\includegraphics[width=0.22\textwidth]{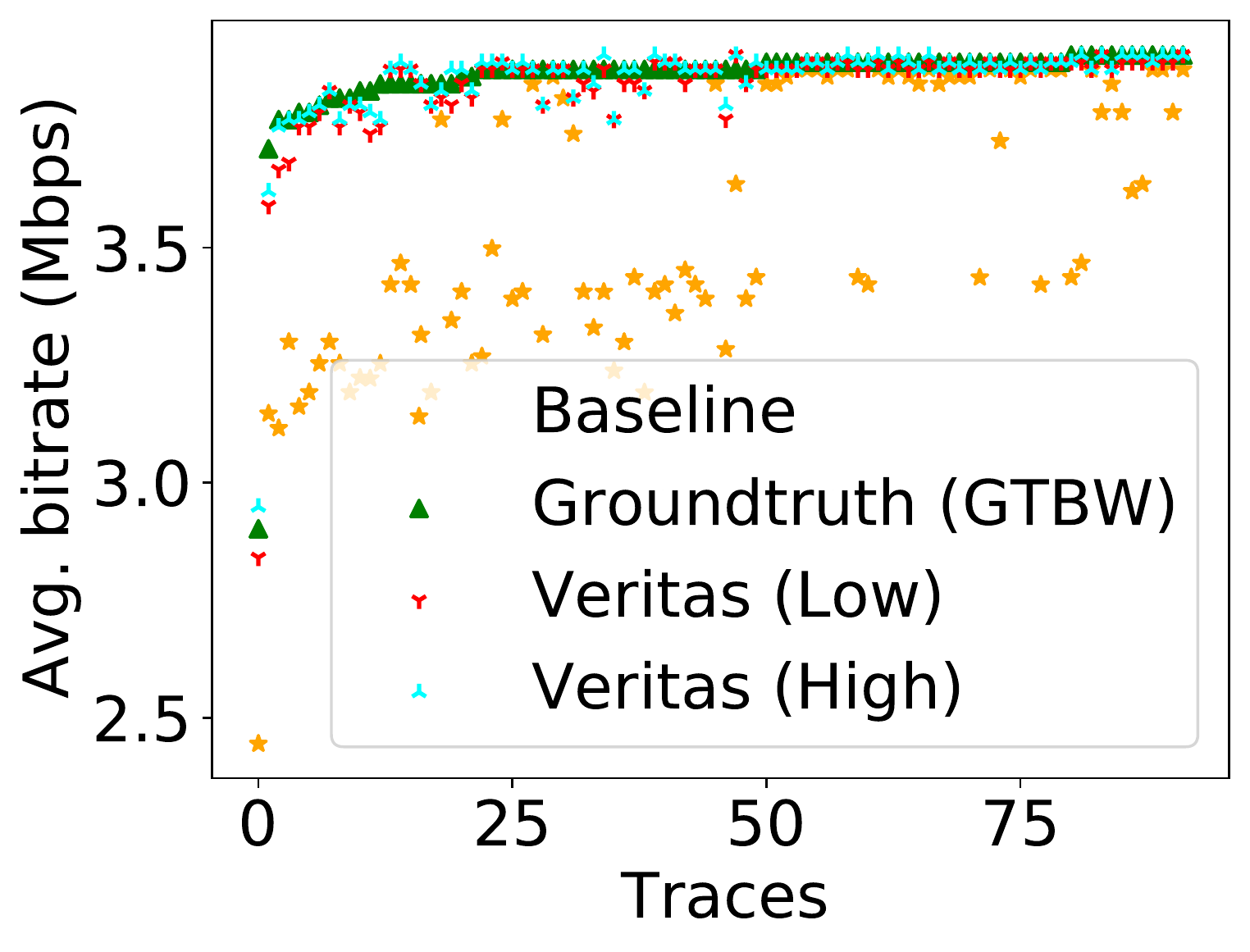}
\label{fig:appendix-bitrate-buffer}
}
\subfigure[Avg. bitrate for changing qualities.]{
\includegraphics[width=0.22\textwidth]{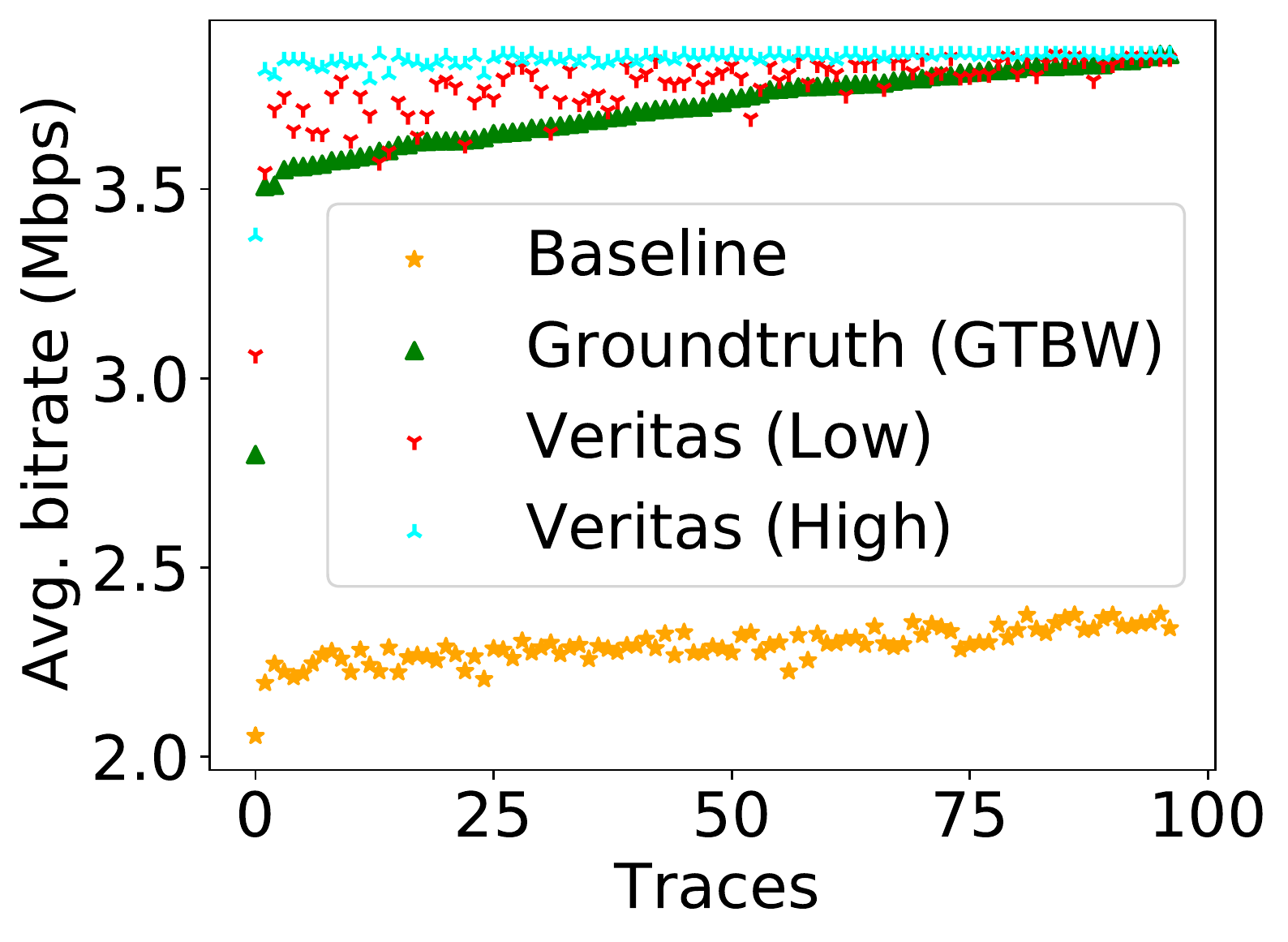}
\label{fig:appendix-bitrate-quality}
}
\caption{Avg. bitrate for counterfactual queries.}
\label{fig:appendix-bitrate}
\end{figure}

\noindent
~\Cref{fig:appendix-bitrate} compares the Avg. bitrate for Baseline and ~\System{} with \Capacity{} for various counterfactual queries.

\subsection{Model and Algorithms}
\label{app:algs}
In this part, we will clarify some details of our models and present pseudo code for all algorithms: Viterbi variant, Baum-Welch forward-backward variant and network throughput estimator used in our EHMM.

\paragraph{Why $B_{s_{1:N}}$ need not be observed.} In \Cref{fig:design-cdg}'s DAG, start time $s_{1:N}$ are not defined as random variables to simplify exposition. If we had defined $s_{1:N}$ as an observed random variable, $s_{1:N}$ could have been used in place of $B_{s_{n-1}}$ to define the sufficient set of observed variables in our $d$-separation argument.
Looking at the dependence between $P\big( C_{s_n} = j\epsilon \big| C_{s_{n - 1}} = i\epsilon \big)$ and $\Delta_{n}$ makes it clear that observing $s_{1:N}$ is also necessary for our Markov model. To conclude, then, we do not actually need to log $B_{s_{1:N}}$ since $s_{1:N}$ is necessary and sufficient and readily available in the trace.

\paragraph{Algorithm Pseudo Codes.}
As introduced in \Cref{sec:abduction}, our Viterbi and Baum-Welch forward-backward variants are nearly the same as their origins, but replace the transition matrix from constant matrix $A$ to $A^{\Delta_n}$ where $\Delta_n$ is as shown in \Cref{sec:abduction} and \Cref{fig:chunk_to_hmm}, and replace the emission process by our proposal as \Cref{eqn:eval_chunk_cap}.
The pseudo codes of both algorithms are provided in \Cref{alg:viterbi} and \Cref{alg:forward_backward}.

\begin{algorithm}[!htb]
\caption{{\bf Forward-Backward Algorithm.} It first computes forward distribution $\alpha_{n, i} = P\big( C_{s_n} = i\epsilon \big| Y_{1:n}, W_{s_{1:n}}, S_{1:n} \big)$; then computes backward distribution $\beta_{n, i} = P\big( C_{s_n} = i\epsilon \big| Y_{n + 1:N}, W_{s_{n + 1:N}}, S_{n + 1:N} \big)$; and finally achieve conditional joint distribution $\Gamma_{i, j, n} = P\big( C_{s_n} = i\epsilon, C_{s_{n + 1}} = j\epsilon \big| Y_{1:N}, W_{s_{1:N}}, S_{1:N} \big)$ by combining $\alpha$ and $\beta$ for all $i, j$ in \capacity state space from $1$ to $N - 1$ chunks.}
\label{alg:forward_backward}
\KwIn{State Space $\mathcal{C}$, Transition times $T$, Initial distribution $u_1$, Transition matrix $A$, Emission process $E$ (\Cref{eqn:eval_chunk_cap}), Throughputs $Y_{1:N}$, TCP states $W_{s_{1:N}}$, Chunk sizes $S_{1:T}$, interval gaps $\Delta$, capacity unit $\epsilon$}
\KwOut{Conditional Joint Distribution $\Gamma$}
\Comment{Alias}
$\xi^\text{back}_{i, j, n} = A^{\Delta_n}_{i, j} E\big( Y_{n}, W_{s_n}, S_{n} \big| j\epsilon \big), \forall i, j \in \mathcal{C}, \forall 2 \leq n \leq N$ \\
$\xi^\text{fore}_{i, j, n} = A^{\Delta_{n + 1}}_{i, j} E\big( Y_{n + 1}, W_{s_{n + 1}}, S_{n + 1} \big| j\epsilon \big), \forall i, j \in \mathcal{C}, \forall 1 \leq n \leq N - 1$ \\
\Comment{Forward}
$\alpha_{1, i} = u_{1, i} E\big( Y_{1}, W_{s_1}, S_{1} \big| i\epsilon \big), \forall i \in \mathcal{C}$ \\
\For{$n = 2 \longrightarrow N$}{
    $\alpha_{n, i} = \sum\limits_{j \in \mathcal{C}} \alpha_{n - 1, j} \xi^\text{back}_{j, i, n}, \forall i \in \mathcal{C}$ \\
}

\Comment{Backward}
$\beta_{N, i} = 1, \forall i \in \mathcal{C}$ \\
\For{$n = N - 1 \longrightarrow 1$}{
    $\beta_{n, i} = \sum\limits_{j \in \mathcal{C}} \xi^\text{fore}_{i, j, n} \beta_{n + 1, j}, \forall i \in \mathcal{C}$ \\
}

\Comment{Posterior}
\For{$n = 1 \longrightarrow N - 1$}{
    \For{$i \in \mathcal{C}$}{
        \For{$i \in \mathcal{C}$}{
            $\Gamma_{i, j, n} = \frac{\alpha_{n, i} \xi^\text{fore}_{i, j, n} \beta_{n + 1, j} \beta_{n + 1, j}}{\sum\limits_{k \in \mathcal{C}} \sum\limits_{l \in \mathcal{C}} \alpha_{t, k} \xi^\text{fore}_{k, l, n} \beta_{n + 1, l} \beta_{n + 1, l}}$ \\
        }
    }
}
\end{algorithm}

\begin{algorithm}[h]
\caption{{\bf Viterbi Algorithm.} It can search for the most likely \capacity state trace $I^{*}$ which can generate given observations $Y_{1:N}, W_{s_{1:N}}, S_{1:N}$, on all chunks through dynamic programming.}
\label{alg:viterbi}
\KwIn{State Space $\mathcal{C}$, Transition times $T$, Initial distribution $u_1$, Transition matrix $A$, Emission process $E$ (\Cref{eqn:eval_chunk_cap}), Throughputs $Y_{1:N}$, TCP states $W_{s_{1:N}}$, Chunk sizes $S_{1:T}$, interval gaps $\Delta$, capacity unit $\epsilon$}
\KwOut{Most Likely State Trace $I^{*}$}
$\xi_{1, i} = u_{1, i} E\big( Y_{1}, W_{s_1}, S_{1} \big| i\epsilon \big), \forall i \in \mathcal{C}$ \\
\For{$n = 2 \longrightarrow N$}{
    $x_{n, i} = \mathop{\arg\max}\limits_{j \in \mathcal{C}} \xi_{1, j} A^{\Delta_n}_{j, i} E\big( Y_{n}, W_{s_n}, S_{n} \big| i\epsilon \big), \forall i \in \mathcal{C}$ \\
    $\xi_{n, i} = \xi_{1, x_{n, i}} A^{\Delta_n}_{x_{n, i}, i} E\big( Y_{n}, W_{s_n}, S_{n} \big| i\epsilon \big), \forall i \in \mathcal{C}$ \\
}
$I^{*}_N = \mathop{\arg\max}\limits_{i \in \mathcal{C}} \xi_{N, i}$ \\
\For{$n = N - 1 \longrightarrow 1$}{
    $I^{*}_n = x_{n + 1, I^{*}_{n + 1}}$ \\
}
\end{algorithm}

\begin{algorithm}[b]
\caption{Network throughput estimator:~\ModelF}
\label{alg:model_f}
\KwIn{$C$, TCP state $W_{S_n}$, Chunk size $S_n$}
\KwOut{$Y_n$}
\Comment{Calculating ssthresh and cwnd.}
\If{$W_{S_n}^{last\_snd} > W_{S_n}^{rto}$} {
    \Comment{Slow start restart.}
    $init\_cwnd \gets 10$
    
    \While{(($W_{S_n}^{last\_snd} -  W_{S_n}^{rto}) > 0$) and 
    ($W_{S_n}^{cwnd} > init\_cwnd))$}{
        $W_{S_n}^{last\_snd} = W_{S_n}^{last\_snd} - W_{S_n}^{rto}$ \\
        $W_{S_n}^{cwnd} \gets W_{S_n}^{cwnd} << 2$
    } 
    $W_{S_n}^{ssthresh} \gets max(W_{S_n}^{ssthresh}, (W_{S_n}^{cwnd} >> 1) + (W_{S_n}^{cwnd} >> 2))$
}
\Comment{Get number of data segments.}
$data\_segments \gets get\_segments(S_n)$ \\
$bdp\_segments \gets get\_segments(\Capacity{}*W_{S_n}^{min\_rtt})$

\eIf{$W_{S_n}^{cwnd} > bdp\_segments$}{
    \eIf{$data\_segments > bdp\_segments$}{
    \Return $C$
    } {
    \Return $S_n / W_{S_n}^{min\_rtt}$
    }
} {
    $rounds \gets 0$ \\
    $sent \gets 0$\\
    \While {$sent < data\_segments$} {
    $sent \gets sent + min(W_{S_n}^{cwnd}, bdp\_segments)$ \\
    \eIf{$W_{S_n}^{cwnd} < W_{S_n}^{ssthresh}$}{
        $W_{S_n}^{cwnd} \gets 2 * W_{S_n}^{cwnd}$ \\
    }{
        $W_{S_n}^{cwnd} \gets W_{S_n}^{cwnd} + 1$
    }
    $rounds \gets  rounds + 1$ \\
    }
    \Return $min((S_n/ (rounds * W_{S_n}^{min\_rtt}), C)$

}
\end{algorithm}

We use a simple model \ModelF, which estimates throughput given \capacity, TCP state and size of related download chunk.
The pseudo code is provided in \Cref{alg:model_f}.



\end{document}